\documentclass[aps,pra,notitlepage,onecolumn,amsmath,amssymb,showpacs,showkeys,nobibnotes]{revtex4-1}
\usepackage{graphicx}
\usepackage{amsmath}
\usepackage{bm}
\usepackage{cleveref}

\newcommand{\be}{\begin{equation}}  
\newcommand{\ee}{\end{equation}}
\newcommand{\ba}{\begin{array}}
\newcommand{\ea}{\end{array}}
\newcommand{\bea}{\begin{eqnarray}}
\newcommand{\eea}{\end{eqnarray}}

\newcommand{\bra}{\langle}
\newcommand{\ket}{\rangle}

\newcommand{\nn}{\nonumber}

\begin{document}

\title{Quantum coherence, many-body correlations, and non-thermal effects for autonomous thermal machines}

\author{C.L. Latune$^{1}$}
\email[correspondence to: ]{lombardlatunec@ukzn.ac.za}
\author{I. Sinayskiy$^{1}$, F. Petruccione$^{1,2,3}$}
\affiliation{$^1$Quantum Research Group, School of Chemistry and Physics, University of
KwaZulu-Natal, Durban, KwaZulu-Natal, 4001, South Africa\\
$^2$National Institute for Theoretical Physics (NITheP), KwaZulu-Natal, 4001, South Africa\\
$^3$School of Electrical Engineering, KAIST, Daejeon, 34141, Republic of Korea}

%


\date{\today}
\begin{abstract}
One of the principal objectives of quantum thermodynamics is to explore quantum effects and their potential beneficial role in thermodynamic tasks like work extraction or refrigeration. So far, even though several papers have already shown that quantum effect could indeed bring quantum advantages, a global and deeper understanding is still lacking.
 Here, we extend previous models of autonomous machines to include quantum batteries made of arbitrary systems of discrete spectrum.  
We establish their actual efficiency, which allows us to derive an efficiency upper bound, called maximal achievable efficiency, shown to be always achievable, in contrast with previous upper bounds based only on the Second Law. Such maximal achievable efficiency can be expressed simply in term of the {\it apparent temperature} of the quantum battery. This important result appears to be a powerful tool to understand how quantum features like coherence but also many-body correlations and non-thermal population distribution can be harnessed to increase the efficiency of thermal machines.

 \end{abstract}

\maketitle

\begin{figure}[ht]
\centering
\includegraphics[width=17.5cm, height=2.8cm]{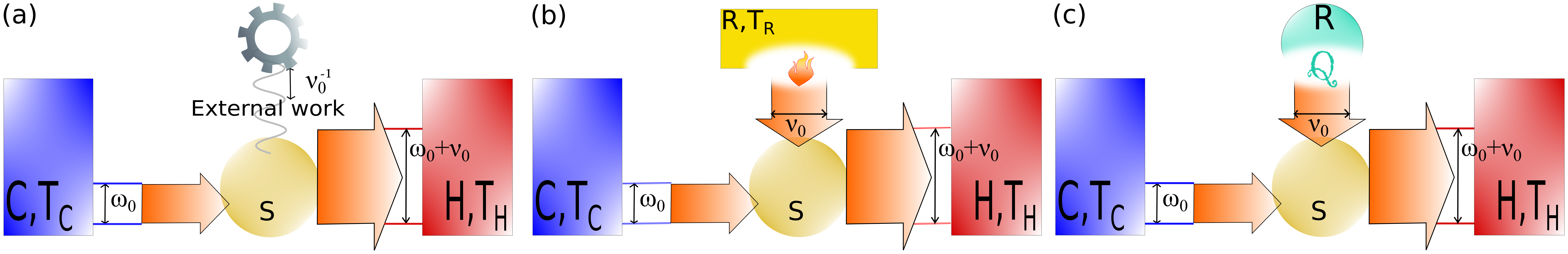}
\caption{Quantum refrigerators composed of a working medium $S$ interacting resonantly at $\omega_0$ with the cold bath $C$. The hot bath $H$, resonant at frequency $\omega_0+\nu_0$, receives quanta of energy $\omega_0+\nu_0$ ($\hbar=1$) from $S$ thanks to an energy input $\nu_0$ from (a) an external source, or (b) a thermal bath $R$ at temperature $T_R$, or (c) a quantum battery $R$ boosted by quantum effects. An engine (promoting energy extraction) is obtained by reverting the energy flows.}
\label{thermalmachines}
\end{figure}

Quantum machines aim to attend to technological and experimental needs \cite{mohammady_low-control_2018,silveri_theory_2017}
 of nano-scale non-invasive devices capable of cooling or loading energy in single quantum systems (e.g. nano-resonators, cantilevers, atoms). A parallel objective is to explore to which extent genuine quantum effects can assist or enhance the performance of such machines like they do for quantum computation and quantum metrology. 
Some of the most notorious quantum and non-equilibrium characteristics, quantum coherence and correlations, where shown in \cite{latune_apparent_2018} to turn quantum thermodynamics intrinsically different from its classical counterpart. It is therefore essential to investigate what is their impact on quantum machines.
 Following this objective, several papers on cyclic machines investigate the effect of quantum coherence \cite{scully_extracting_2003,scully_extracting_2002,brandner_universal_2017,mehta_quantum_2017,turkpence_quantum_2016,turkpence_photonic_2017}, many-body correlations \cite{zhang_four-level_2007, wang_thermal_2009, dillenschneider_energetics_2009, hardal_superradiant_2015,altintas_quantum_2014,altintas_rabi_2015,hardal_phase-space_2018,doyeux_quantum_2016,li_quantum_2014,jaramillo_quantum_2016,mueller_correlating_2017,turkpence_photonic_2017}, and other non-thermal characteristics (mainly squeezing) \cite{gardas_thermodynamic_2015,leggio_otto_2016,abah_efficiency_2014,rosnagel_nanoscale_2014,manzano_entropy_2016, huang_effects_2012} on thermodynamic tasks (refrigeration or work/energy extraction). Similarly, studies investigated the effects of coherence \cite{niedenzu_performance_2015} and correlations \cite{gelbwaser-klimovsky_power_2015} on semi-classical continuous machines (simultaneous and continuous interaction with both the cold and hot baths) driven by external controls (Fig. \ref{thermalmachines}, panel a).
 
However, such cyclic or semi-classical machines require time-dependent external controls, raising questions regarding the energetic cost of such operations \cite{clivaz_unifying_2017} and doubts about the overall energetic balance of the machines. Moreover, contacts with classical systems (necessary for external controls) make them not well-fitted for non-invasive and local applications. 
Above all, the study of quantum effects in such machines often requires   
 baths in very unprobable non-thermal states (and very demanding energetically and experimentally to prepare). 
 A more viable alternative to baths is offered by the collisional model. Nevertheless, its requirement of short and repeated interactions together with a high number of reinitializations or preparations of identical systems also represents experimental difficulties.

One promising alternative avoiding the above drawbacks is autonomous machines.  
 Differently from cyclic and semi-classical machines (shown to be equivalent \cite{uzdin_equivalence_2015}), autonomous machines operate autonomously with no need of external work or controls (which ensures that all energetic and entropic contributions are taken into account). The absence of external controls makes them more suitable for nano-scale and non-invasive operations. 
Most studies focus on autonomous machines involving three thermal baths, sometimes called absorption refrigerators (Fig. \ref{thermalmachines}, panel b), where one of the bath plays the role of the energy source of the machine \cite{brunner_entanglement_2014,brask_small_2015,mitchison_coherence-assisted_2015,palao_quantum_2001, levy_quantum_2012,venturelli_minimal_2013,correa_performance_2013,correa_quantum-enhanced_2014,correa_optimal_2014,correa_multistage_2014,hofer_autonomous_2016,mitchison_realising_2016,he_enabling_2017,mari_cooling_2012,linden_how_2010, skrzypczyk_smallest_2011, kosloff_quantum_2014,maslennikov_quantum_2017, du_nonequilibrium_2018}.

Nevertheless, there is an other type of autonomous machines whose source of energy comes from a quantum battery, namely an ancillary system of finite size (i.e. not a bath)(Fig. \ref{thermalmachines}, panel c). The quantum battery interacts continuously with the machine, offering naturally a platform more adapted and convenient than baths to explore quantum and non-thermal effects. Such machines are more challenging theoretically than semi-classical machines or than absorption refrigerators due to the continuous interaction with the quantum battery which cannot be treated as a bath. 
 Methods used for cyclic machines cannot be applied for these autonomous machines. Instead, one has to establish the dissipative dynamics of the working media together with the quantum battery, which becomes a particularly complex task since their total Hamiltonian is not necessarily diagonalisable (due to their interaction).  
Such machines received little attention so far. Indeed, it is still unclear whether quantum and non-thermal effects can bring advantages to autonomous thermal machines.
 Previous studies, pioneered in \cite{tonner_autonomous_2005} and continued in \cite{boukobza_breaking_2013, gelbwaser-klimovsky_heat-machine_2014,gelbwaser-klimovsky_work_2015} already started to address this  problem and provided a general upper bound. However, such upper bound gives little  information about the specific role of coherence and correlations. Moreover, its attainability is not certain and not discussed.

Here, we study in details the impact of quantum and non-thermal features on the performance of autonomous thermal machines in a broadly extended framework. 
 After establishing the actual efficiency we provide the maximal achievable efficiency.
 Interestingly, it can be expressed simply in term of the concept of apparent temperature introduced in \cite{latune_apparent_2018}. 
 This allows us to investigate straightforwardly the impact on autonomous thermal machines of three non-thermal features: quantum coherence, many-body correlations, and non-thermal population distribution.

\subsection*{Results}
{\bf The model}. The aim of a thermal machine is to reverse the natural heat flow between two thermal baths $C$ and $H$ of different temperature $T_C$ and $T_H$, respectively, or to extract work (or energy) from them. This is achieved by introducing a system $S$ interacting with both $C$ and $H$. Even though we could realise refrigeration or energy extraction with only one single system, traditionally $S$ is used as a connection between the baths and a quantum battery (an ancillary system) which provides or stores energy. We denotes such quantum battery by $R$. The connecting system $S$ is often called working medium. For semi-classical or cyclic machines, $R$ is not included in the physical description and thus behaves as a classical system. By contrast, in autonomous thermal machines, $R$ is included in the physical description and the ensemble $SRCH$ is assumed to evolve unitarily through a {\it time-independent} Hamiltonian (ensuring no external source of work). Although no coupling is considered between $R$ and the baths, they end up interacting indirectly (through $S$). The global Hamiltonian is $H_{global}= H_S + H_R + H_B + V_{SR} + V_{SB}$,
where $H_X$, $X=S,R,B$ are free Hamiltonians of the corresponding subsystems and $B$ collectively denotes the two baths $C$ and $H$. $V_{SR}$ ($V_{SB}$) is the coupling Hamiltonian between $S$ and $R$ ($B$). In the following we regroup the terms $H_S$, $H_R$ and $V_{SR}$ under the notation $H_{SR}:=H_S+H_R+V_{SR}$.

\begin{figure}[ht]
\centering
\includegraphics[width=0.45\textwidth]{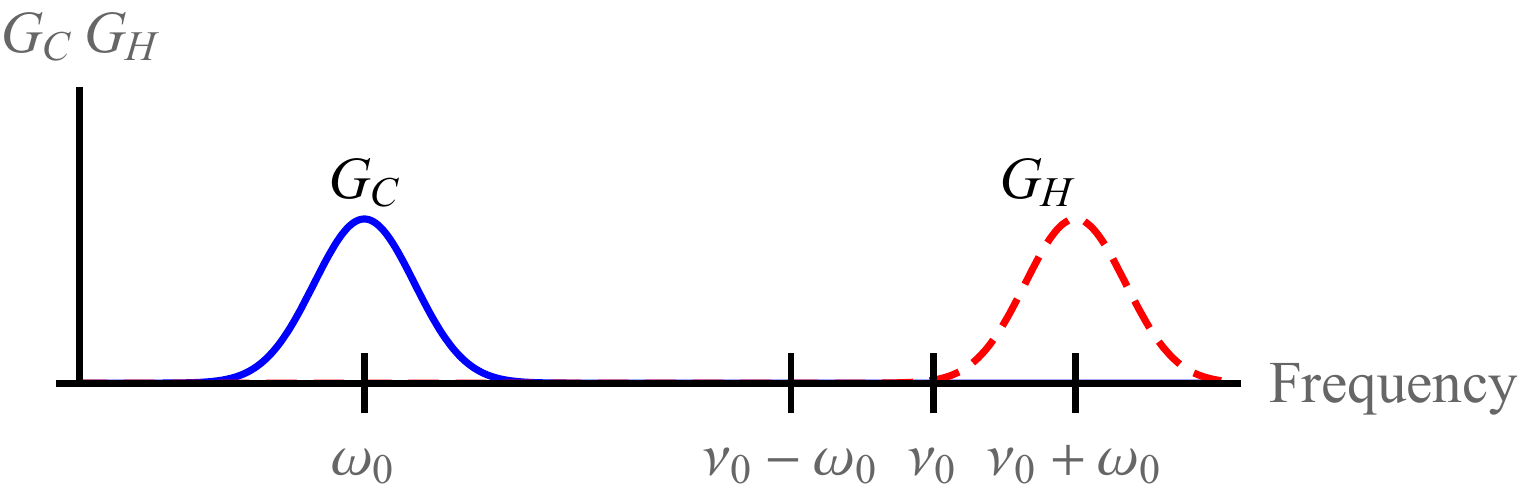}
\caption{Bath spectral densities $G_C$ and $G_H$ of the baths $C$ and $H$, respectively. The cold bath $C$ is resonant with the transition energy $\omega_0$ of $S$, whereas the hot bath $H$ is resonant with the transition $\omega_0+\nu_0$ of $SR$. 
}\label{fig2}
\end{figure}

From the point of view of the Second Law, expressed and discussed in Methods (`Upper bound from the Second Law'), it might seem that one just needs to inject energy and/or entropy in order to realise a thermal machine. However, it is no so simple. One has to design a device whose dynamics actually inverts the natural heat flow. The second law only provides limits on the performance, but gives no clue about how to realise such machines. 
Our model is an extension of the one introduced in \cite{gelbwaser-klimovsky_heat-machine_2014,gelbwaser-klimovsky_work_2015}.
One of the key feature is that $S$ and $R$ are dispersively coupled through an Hamiltonian of the form $V_{SR} = g N_S A_R$ ($\hbar =1$) where $g$ characterises the strength of the coupling, $A_R$ is an observable of $R$, and $N_S = \alpha H_S$ with $\alpha$ a positive constant. We justify the use of a dispersive coupling in Methods (`Why dispersive coupling?') showing that it seems to be the only universal coupling allowing for refrigeration or energy extraction for any working media.  
 In order to avoid heat leaks and optimise the efficiency the working medium $S$ has to be resonant with only one of the two baths (otherwise energy will flow directly through $S$ from the hot bath to the cold). The other bath has to be resonant with $SR$ in order to get $R$ involved in the dynamics.
 Thermal machines with multiple resonances with the baths are possible although more complex, therefore we focus on the simpler design with only one resonant transition per bath. Then, we assume that $C$ is resonant with $S$, denoting by $\omega_0$ the corresponding transition frequency, and $H$ is resonant with $SR$ at the frequency $\omega_0+\nu_0$ (see Fig. \ref{fig2}), where $\nu_0$ is a transition frequency of $R$ (the choice $H$ resonant with $\omega_0$ and $C$ with $\omega_0 -\nu_0$ yields equivalent dynamics). 
 
 One should note that $R$ can have other transitions. If so, the width of the bath spectral density $G_H$ ($G_C$) of the hot (cold) bath (defined in \eqref{defgamma} and \eqref{defspectraldensity})
 might need to be reduced in order to avoid resonance with these other transitions. Such baths, if not available directly, can be obtained through filtering or coupling to an intermediary two-level system \cite{geusic_quantum_1967, levy_quantum_2012, kosloff_quantum_2014, correa_optimal_2014, gelbwaser-klimovsky_heat-machine_2014,gelbwaser-klimovsky_work_2015, mitchison_realising_2016}.
   Consequently, instead of a harmonic oscillator as quantum battery (as in \cite{boukobza_breaking_2013, gelbwaser-klimovsky_heat-machine_2014,gelbwaser-klimovsky_work_2015}) we can consider a large class of systems, namely arbitrary system of discrete spectrum (such that the resonance conditions can be satisfied through bath engineering when necessary).  
 In principle, the same is valid for $S$. However, in order to simplify the derivation of the main results we restrict $S$ to be either a two-level system (as in \cite{gelbwaser-klimovsky_heat-machine_2014,gelbwaser-klimovsky_work_2015}) or a harmonic oscillator, which provides more flexibility and possibilities for experimental realisations.  

The coupling with the baths are considered of the following general form $V_{SB} =  A_S A_B$, where $A_S$, $A_B = \lambda_H A_{H} + \lambda_C  A_C$ are observables of $S$, $H$, and $C$ respectively, and the constant $\lambda_H$ ($\lambda_C$) characterises the strength of the coupling between $S$ and $H$ ($C$). In the following, such constant are included in $A_C$ and $A_H$, respectively. \\

{\bf Dynamics}. Assuming a weak coupling with the baths ($\lambda_H$ and $\lambda_C$ much smaller than the inverse of the bath correlation times) we use the Born-Markov approximation and the formalism of \cite{breuer_theory_2007} to derive corrections to the local approach \cite{trushechkin_perturbative_2016} of the master equation of $SR$. The exact global approach \cite{levy_local_2014,gonzalez_testing_2017} is intractable in general. The main steps of the derivation are described in Methods `Expression of the baths' dissipative operators'. 
The effects of $V_{SR}$ are taken into account up to second order in $g/\nu_0$ for weak coupling $g \ll \nu_0$ and $g\ll \lambda_C, \lambda_H$. 
Under such conditions $R$ evolves slowly through the (indirect) influence of the bath $H$, at a timescale $\tau_R:= \left (G_C(\omega_0) \frac{g^2}{|\nu|^2}\right )^{-1}$. $\tau_R$ is much larger than the timescale $\tau_{es}:= [G_C(\omega_0)]^{-1}$ under which $S$ is brought to a quasi steady state (thermal state at temperature $T_C$ with corrections of order $g/\nu_0$) due to its resonant interaction with $C$. Moreover, defining the internal energy of the subsystem $X = R, S$, or $RS$ as $E_X :=  \bra H_X\ket_{\rho_{SR}}$ (using the notation $\bra {\cal O}\ket_\rho $ to denote ${\rm Tr}\rho {\cal O}$ the expectation value of the operator ${\cal O}$ in the state $\rho$), one can show that $\dot{E}_S = {\cal O}(g^3/\nu_0^3)$ for times much larger than $\tau_{es}$. It means that $S$ can be regarded as ``energetically in steady state'' 
so that the energy (in form of quanta $\nu_0$) slowly provided by $R$ is transferred to $H$ together with energy taken from $C$ (in form of quanta $\omega_0$) in order to satisfy the resonance condition of $H$ ($\omega_0+\nu_0$). This leads to the refrigeration of $C$ and happens only if the condition \eqref{refricond1} is fulfilled. Eventually, after long times (much larger than $\tau_R$) $R$ reaches a steady state and the machine stops working. This is the price to pay for using finite size quantum batteries.
 In the remainder of the paper we consider the above regime of $t \gg \tau_{es}$. 

The heat flow from the bath $j=H,C$ to the ensemble $SR$ is defined as \cite{alicki_quantum_1979}
 \be\label{defhflow}
 \dot{Q}_{SR/j}:= {\rm Tr}_{SR}{\cal L}_j \rho_{SR}^I H_{SR},
 \ee
where $\rho_{SR}^I$ denotes the density matrix in the interaction picture with respect to $H_{SR}$, and the dissipator ${\cal L}_j$ corresponds to the action of the bath $j=H,C$ on $SR$. According to the above definitions the following equality holds (First Law) $\dot{E}_{SR} = \dot{Q}_{SR/H} + \dot{Q}_{SR/C}$.
For times much larger than $\tau_{es}$, the heat flow leaving the cold bath, obtained by introducing the expression of ${\cal L}_C$ in \eqref{defhflow}, is (see details in Methods and Supplementary Information) 
\bea\label{mtdotqc}
\dot{Q}_{SR/C} &\propto& \left({g^2\over\nu_0^2}\right) \omega_0 \Big(e^{-{\omega_0\over T_C}} \langle A_R^{\dag}(\nu_0)A_R(\nu_0) \rangle_{\rho_{R}^I} - e^{-{(\omega_0+\nu_0)\over T_H}}\langle A_R(\nu_0)A_R^{\dag}(\nu_0)\rangle_{\rho_R^I}\Big) + {\cal O}\left({g^3\over\nu_0^3}\right).
\eea 
 The above expression is a generalisation of the one obtained in \cite{gelbwaser-klimovsky_heat-machine_2014, gelbwaser-klimovsky_work_2015,ghosh_catalysis_2017} for harmonic oscillators (the Boltzmann constant is set equal to $1$). $A_R(\nu_0)$ and $A_R^{\dag} (\nu_0)$ are the eigenoperators for the transition energy $\nu_0$, defined by 
$A_R(\nu_0) = \sum_{\epsilon' -\epsilon = \nu_0}\Pi(\epsilon)A_R\Pi(\epsilon')$ (and the hermitian conjugate for $A_R^{\dag} (\nu_0)$), with $\Pi(\epsilon)$ being the projector onto the eigenspace of $H_R$ associated to the eigenenergy $\epsilon$. 
 Moreover, the following relations hold for $t \gg \tau_{es}$,
\be\label{mtsseq}
{\dot{Q}_{SR/C}\over \omega_0}=-{\dot{Q}_{SR/H}\over\omega_0+\nu_0}+{\cal O}\left({g^3\over\nu_0^3}\right)=-{\dot{E}_R\over \nu_0}+{\cal O}\left({g^3\over\nu_0^3}\right) .
\ee

{\bf Actual (universal) efficiency}. 
We focus firstly on refrigeration. The energy extraction regime will be treated in a second time.
The efficiency refrigeration $\eta_r$ 
is defined as the ratio of the heat extracted from $C$, accounted by $\dot{Q}_{SR/C}$, by the energy invested (and provided by $R$), accounted by $-\dot{E}_R $, $\eta_r := { \dot{Q}_{SR/C} \over -\dot{E}_R}$.
From the above relation \eqref{mtsseq} we deduce that 
\be\label{eta}
\eta_r = \frac{\omega_0}{\nu_0} + {\cal O}(g^3/\nu_0^3).
\ee

Remarkably, the efficiency is constant and independent from the initial state of $R$. 
This is contrasting with the intuition created by the Second Law and its associated upper bound \eqref{2dlawupperbound}. Eq. \eqref{eta} extends to quantum batteries of discrete spectrum in any state the result already known for thermal two-level systems or thermal baths \cite{skrzypczyk_smallest_2011, correa_quantum-enhanced_2014, geusic_quantum_1967}. However, one should note that even though the actual efficiency does not depend on the initial state of $R$, the power of the machine does depend on it, but also the operating regime (refrigeration or energy extraction). 
Alternatively, for a given state of $R$, one can ask what is the maximum achievable efficiency (adjusting parameters of the setup like the resonance $\omega_0$). This is the object of the next paragraph. \\

{\bf Maximal achievable efficiency}. From \eqref{eta} one can see that in order to increase the efficiency one only needs to increase $\omega_0$. However, if $\omega_0$ is too large, the refrigerator may stop refrigerating. Then, the value of $\omega_0$ has to be subjected to the constraint that \eqref{mtdotqc} should remain positive, $\dot{Q}_{SR/C}\geq0$. According to \eqref{mtsseq} this happens simultaneously with $\dot{E}_R \leq 0$, meaning that $R$ is powering the machine.
 This leads to the following necessary and sufficient condition for refrigeration, 
\be\label{refricond1}
\omega_0 \leq \nu_0 \frac{T_C}{T_H-T_C} \left(1 -\frac{T_H}{{\cal T}_R}\right),
\ee
where ${\cal T}_R$ is the {\it apparent temperature} of $R$, defined as
\be\label{mtapptemp}
{\cal T}_R:=  \nu_0\left( \ln{\frac{\langle A_R(\nu_0)A_R^{\dag}(\nu_0)\rangle_{\rho_R^I}}{\langle A_R^{\dag}(\nu_0)A_R(\nu_0)\rangle_{\rho_R^I}}}\right)^{-1},
\ee
and introduced in \cite{latune_apparent_2018}. The above Eq. \eqref{refricond1} leads straightforwardly to the {\it maximal achievable efficiency} $\eta_{\rm ac}$ (achieved at zero power, as usual),
\be\label{effupperbound}
\eta_r={\omega_0\over\nu_0} + {\cal O}\left({g^3\over\omega_0^3}\right) \leq \eta_{\rm ac}:= \frac{T_C}{T_H-T_C} \left(1 -\frac{T_H}{{\cal T}_R}\right).
\ee

The apparent temperature was shown to determine the heat flows between out-of-equilibrium quantum systems and general reservoir (bath or collisional model) \cite{latune_apparent_2018}, appearing as a quantifier of a system's tendency to cede packets of quantised energy. 
For thermal states it coincides with the usual temperature (appearing in the Boltzmann distribution). It is therefore remarkable that the maximal achievable efficiency can be expressed in term of the apparent temperature of $R$. This brings a physically meaningful upper bound: when $R$ is in a thermal state, the achievable upper bound is given by the usual Carnot bound, and when $R$ is in a non-thermal state, the achievable maximal efficiency is simply given by substituting the temperature by the apparent temperature. Furthermore, this re-enforces and extends the relevance of the concept of apparent temperature beyond its original framework.

Finally, this result is important for two more reasons. First, it provides an {\it achievable} maximal efficiency, 
which is not provided by previous upper bounds derived from the Second Law. The main reason is that the upper bound \eqref{2dlawupperbound} deduced from the second law is expressed in term of $\dot{S}(\rho_R)$, the entropy change rate of $R$. On the other hand, this upper bound is supposed to be reached when the entropy production rate $\dot{\Sigma}_R$ (always positive) is equal to zero (sometimes also called the reversibility condition \cite{niedenzu_quantum_2018}). However, when $\dot{\Sigma}_R$ is equal to zero, $\dot{S}(\rho_R)$ is smaller. In other words, when the upper bound is supposed to be saturated the value of $\dot{S}(\rho_R)$ has changed, bringing doubts whether the upper bound \eqref{2dlawupperbound} can ever be reached. 
Moreover, it is not guaranteed that one can ever engineer a machine such that $\dot{\Sigma}_R = 0$. 
Such problematics regarding upper bounds were recently discussed in \cite{niedenzu_quantum_2018} (for cyclic machines). 
From an other perspective, the upper bound \eqref{2dlawupperbound} gives the misleading idea that the efficiency can be increased by increasing the entropy change rate $\dot{S}(\rho_R)$. This is not true in general. The actual efficiency can indeed be expressed as $\eta_r =  {T_C\over T_H-T_C}\left(1 + T_H{\dot{S}_{\rm fl}(\rho_{R})\over -\dot{E}_R }\right)$, obtained by substituting in \eqref{2dlawupperbound} the entropy change rate $\dot{S}(\rho_R)$ by the flow of entropy $\dot{S}_{\rm fl}(\rho_R):=\dot{S}(\rho_R) - \dot{\Sigma}_R$. Then, increasing $\dot{S}(\rho_R)$ is not a guarantee of increase of $\dot{S}_{\rm fl}(\rho_R)$ and therefore of the efficiency. It has been noted for instance that coherences between non-degenerate levels increase the entropy production but do not affect the flow of entropy \cite{santos_role_2017}.

Secondly, using the framework introduced in \cite{latune_apparent_2018} the result \eqref{effupperbound} provides crucial information on how quantum and non-equilibrium effects can be harnessed to boost quantum thermal machines. This is detailed in the following paragraphs. \\

{\bf Recovering known results}. We first show that the maximum achievable efficiency \eqref{effupperbound} provides the usual results for the known situations. As already mentioned above, when $R$ is in a thermal state, ${\cal T}_R$ coincides with the usual temperature (appearing in the Boltzmann distribution) \cite{latune_apparent_2018} and the usual Carnot bound \cite{palao_quantum_2001, linden_how_2010, levy_quantum_2012, correa_performance_2013, correa_quantum-enhanced_2014, correa_optimal_2014} is recovered from \eqref{effupperbound}. 
In the limit of classical batteries, the ladder operators commute, $[A_R(\nu_0),A_R^{\dag}(\nu_0)]=0$, and then the associated apparent temperature is ${\cal T}_R = +\infty$, implying that $R$ behaves as a pure work reservoir. We recover the well-known observation that classical work reservoirs correspond to infinite-temperature thermal baths \cite{strasberg_quantum_2017}. 
Finally, if $R$ is a harmonic oscillator in a squeezed thermal state, its apparent temperature can be expressed in terms of the temperature of the thermal excitation $T_R$ and the squeezing factor $r$ as (see Supplementay Information), ${\cal T}_R = \nu_0 \left[\ln{\frac{ \tanh^2{r} + e^{\nu_0/T_R} }{ \tanh^2{r}e^{\nu_0/T_R} +1}}\right]^{-1},$
so that when substituting in \eqref{effupperbound} the maximal achievable efficiency is the analogue of the upper bound derived in \cite{correa_quantum-enhanced_2014,huang_effects_2012} for machines powered by squeezed thermal baths.\\

\begin{figure}[ht]
\centering
\includegraphics[width=7cm, height=6cm]{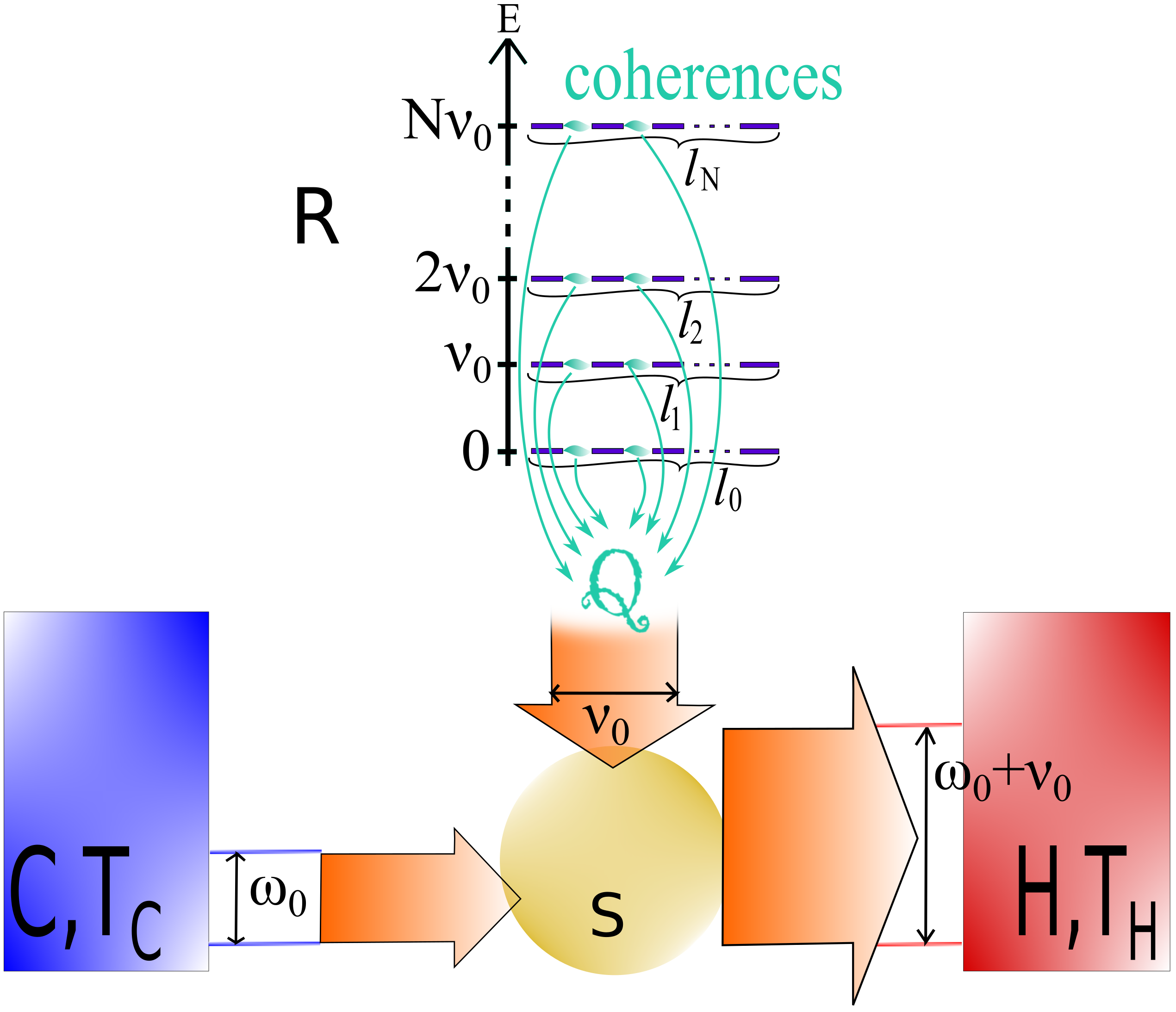}
\caption{Example of a degenerated-level system $R$ used to power an autonomous refrigerator. The energy levels $|n,k\ket$, $n \in [0;N]$, $k\in [1;l_n]$, are represented by purple dashes, with the vertical axis representing the energy of the levels and the horizontal axis the degeneracy. The presence of coherence between degenerated levels (represented by a green link) affect dramatically the apparent temperature. Higher efficiencies are achievable if and only if the coherences satisfy the condition \eqref{cohcond}.}
\label{cohpower}
\end{figure}

{\bf Effects of coherence}. We now combine the framework and results from \cite{latune_apparent_2018} to \eqref{effupperbound}. Only levels taking part in transitions of energy $\nu_0$ contribute to the heat flows (due to the resonance condition). Then, if $R$ has $(N+1)$ energy levels involved in such transitions, we denote them by $|n,k\ket$ with $n \in [0;N]$ and $k$ is the degeneracy index running from $1$ to $l_n \geq 1$, $l_n$ being the number of degeneracy of the level $n$ (see Fig. \ref{cohpower}). In other words, $H_R |n,k\ket = n\nu_0 |n,k\ket$ with $n \in [0;N]$ and $k\in[1;l_n]$.
The ladder operator $A_R(\nu_0)$ can be expressed as (using the expression mentioned after \eqref{mtdotqc}),
\be\label{arsingle}
A_R(\nu_0) = \sum_{n=1}^N \sum_{k = 1}^{l_{n-1}} \sum_{k'=1}^{l_n}\alpha_{n-1,n,k,k'}|n-1,k\rangle  \langle n,k'|,
\ee
with $\alpha_{n-1,n,k,k'} := \langle n-1,k|A_R|n,k'\rangle$. We assume that all transitions amplitudes $\alpha_{n-1,n,k,k'}$ are equal as it does not change the nature of the results and simplify the expressions.
 The situation where $R$ is an infinite-level system (harmonic oscillator) is treated in the following.  
From the definition \eqref{mtapptemp} we find for the apparent temperature \cite{latune_apparent_2018}
\be\label{cohapptemp}
{\cal T}_R = \nu_0 \left( \ln{\frac{\sum_{n=1}^{N} l_n(\rho_{n-1} + c_{n-1})}{\sum_{n=1}^{N} l_{n-1}(\rho_n + c_n)}}\right)^{-1},
\ee
where $\rho_n := \sum_{k=1}^{l_n} \langle n,k|\rho_R|n,k\rangle$ is the sum of the populations of the degenerated levels of energy $n\omega$, 
 and $c_n  := \sum_{k \ne k' \in [1,l_n]} \langle n,k|\rho_R|n,k'\rangle$ 
 is the sum of the coherences between these same degenerated levels.  
 The corresponding maximal achievable efficiency can be re-written as
\be\label{effcoh}
\eta_{\rm ac} =  \frac{T_C}{T_H-T_C} \left[1 -\frac{T_H}{{\cal T}_0} - \frac{T_H}{\nu_0} \ln{\frac{1+{\cal C}^-/\rho^-}{1+{\cal C}^+/\rho^+}}\right],
\ee
where ${\cal T}_0:=\nu_0(\ln \rho^-/\rho^+)^{-1}$ is the apparent temperature of $R$ without coherence (that is all $c_n$ equal to 0), and we defined ${\cal C}^+ := \sum_{n=1}^N l_{n-1}c_n$, $\rho^+:= \sum_{n=1}^N l_{n-1}\rho_n$, ${\cal C}^-:=\sum_{n=1}^N l_{n}c_{n-1}$, and $\rho^-:=\sum_{n=1}^N l_{n}\rho_{n-1}$.

 The result \eqref{effcoh} is important as it provides how coherence affects the efficiency. In particular, it is important to note that coherences between levels of different energy do not affect the apparent temperature and therefore do not confer possibilities of efficiency increase.
The core mechanism relies on the fact that heat exchanges are controlled by the quantities $\bra A_R(\nu_0)A^{\dag}(\nu_0)\ket_{\rho_R}$ and $\bra A_R^{\dag} (\nu_0)A(\nu_0)\ket_{\rho_R}$ determined by the populations but also by coherences between degenerated states of $R$ which ends up affecting the apparent temperature \cite{latune_apparent_2018}  and the achievable efficiency (see Fig \ref{cohpower}).
 The overall balance, which is not always beneficial for the efficiency, is reflected in \eqref{effcoh}. Coherences increase the maximal achievable efficiency if and only if 
 \be\label{cohcond}
  {\cal C}^+\geq {\cal C}^-e^{-\nu_0/{\cal T}_0}.
  \ee
Moreover, ${\cal C}^{\pm}$ can take value close to $\rho^{\pm}$ (with the restriction of positivity and ${\cal C}^{+} + \rho^{+} \geq 0$ and ${\cal C}^{-} + \rho^{-} \geq 0$), which can generate great increase of achievable efficiency.

As an example we mention the phaseonium \cite{scully_extracting_2003,scully_extracting_2002}, which is a three-level system with exited state $|a\ket$ and two (quasi-)degenerated ground states $|b\ket$ and $|c\ket$ (sometimes referred as the $\Lambda$-configuration). For such system, ${\cal C}^+ = 0$, ${\cal C}^- = c_0 = \bra b|\rho_R|c\ket + \bra c|\rho_R|b\ket$, and $\rho^- = \rho_0 =  \bra b|\rho_R|b\ket + \bra c|\rho_R|c\ket$, so that the maximal achievable efficiency would be $\eta_{\rm ac} =  \frac{T_C}{T_H-T_C} \left[1 -\frac{T_H}{{\cal T}_0} - \frac{T_H}{\nu_0} \ln{(1+c_0/\rho_0)}\right].$
 Then, one finds that the coherence in the phaseonium increases the maximal achievable efficiency if and only if $c_0 < 0$, 
 yielding large increases when $c_0$ takes values close to $-\rho_0$.

Conversely, if the system used is a three-level system in the $V$-configuration ($|a\ket$ is the ground state, $|b\ket$ and $|c\ket$ are the degenerated excited states), the maximal achievable efficiency takes the form, $\eta_{\rm ac} =  \frac{T_C}{T_H-T_C} \left[1 -\frac{T_H}{{\cal T}_0} + \frac{T_H}{\nu_0} \ln{(1+c_1/\rho_1)}\right]$, with $c_1 = \bra b|\rho_R|c\ket + \bra c|\rho_R|b\ket$ and $\rho_1 =  \bra b|\rho_R|b\ket + \bra c|\rho_R|c\ket$. The conclusion is opposite to the previous one: coherence increases the maximal achievable efficiency if and only if $c_1 > 0$. \\

\begin{figure}[ht]
\centering
\includegraphics[width=8.5cm, height=4.5cm]{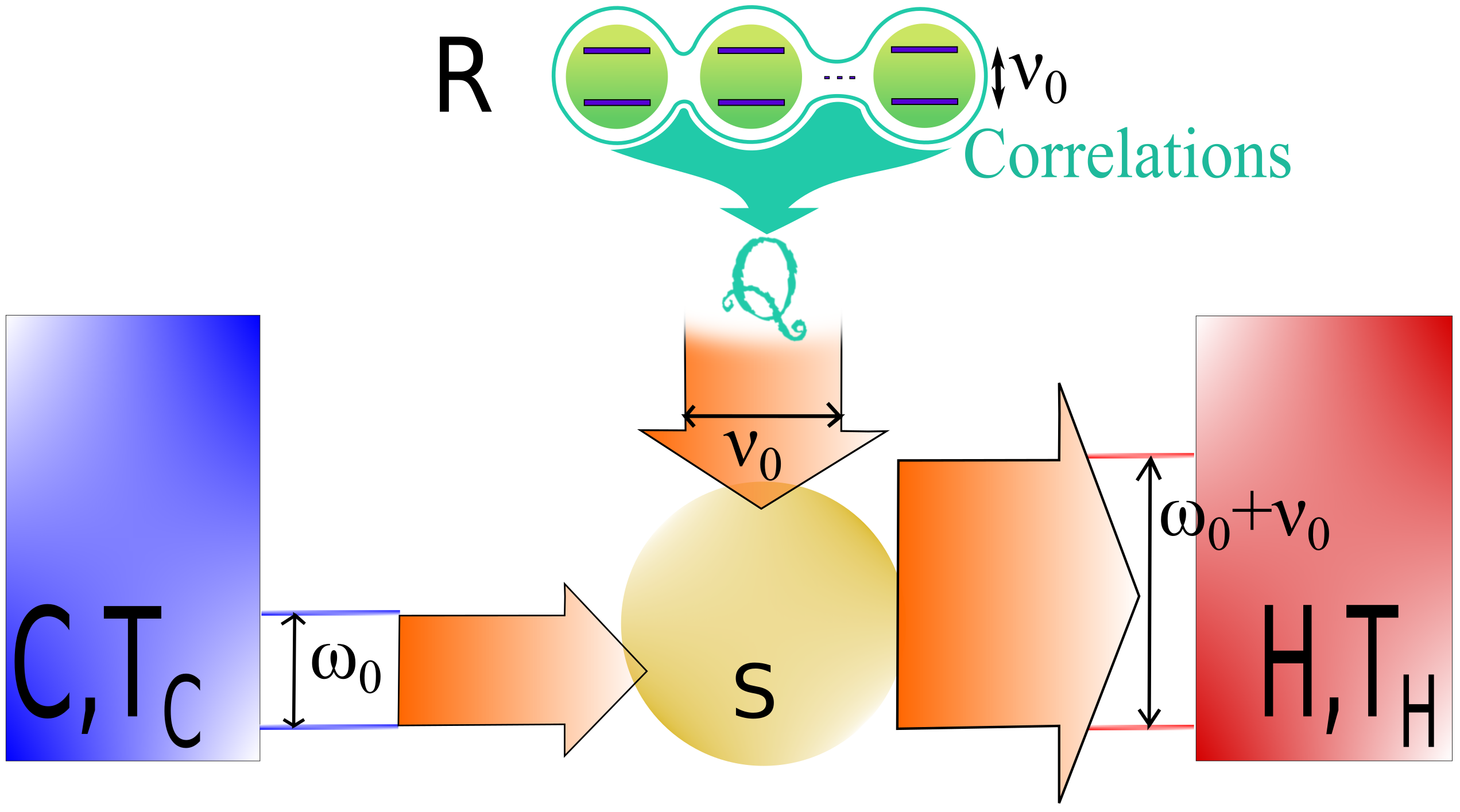}
\caption{Example of a many-body system (here, an ensemble of two-level systems) powering an autonomous refrigerator. The presence of correlations affect dramatically the apparent temperature and higher efficiencies are achievable if and only if the condition \eqref{corrcond} is satisfied.}
\label{refricorr}
\end{figure}

{\bf Effects of correlations}. 
In this paragraph we assume that $R$ is an ensemble of $N$ non-interacting subsystems (not necessarily identical neither with finite number of levels) of {\it same} transition energy $\nu_0$ (Fig. \ref{refricorr}). We require the following important condition that all the $N$ subsystems appear indistinguishable to $S$ (which usually requires confinement in a volume smaller than typical length scales of $S$). 
  Upon the above conditions the $N$ sub-systems interact {\it collectively} with $S$ and the ladder operator $A_R(\nu_0)$ is a {\it collective} ladder operator,
$A_R(\nu_0) = \sum_{i=1}^{N} A_i(\nu_0)$, where $A_i(\nu_0)$ is the ladder operator of the subsystem $i$ (with the same properties as above). 
Then, applying definition \eqref{mtapptemp}, the apparent temperature of the ensemble $R$ is,
\bea\label{correlationtemp}
{\cal T}_R &=&  \nu_0 \left( \ln{\frac{ \sum_{i=1}^m \langle A_i(\nu_0) A_i^{\dag}(\nu_0)\rangle_{\rho_R}+c }{\sum_{i=1}^m \langle A_i^{\dag}(\nu_0) A_i(\nu_0) \rangle_{\rho_R} + c}}\right)^{-1},
\eea
where $c$ is the sum of the correlations (and product of local coherences, see \cite{latune_apparent_2018}), $c:=\sum_{i\ne j=1}^m \langle A_i(\nu_0) A_j^{\dag}(\nu_0)\rangle_{\rho_R} = \sum_{i\ne j=1}^m \langle A_i^{\dag}(\nu_0) A_j(\nu_0)\rangle_{\rho_R}$. 
As previously with coherence, {\it correlations act as populations}, and their impact on the apparent temperature can be significant, enabling high efficiency increases. 
The maximal achievable efficiency is
\be\label{effcorr}
\eta_{\rm ac} = \frac{T_C}{T_H-T_C} \left[1 -\frac{T_H}{{\cal T}_0} - \frac{T_H}{\nu_0} \ln{\frac{1+c/n^-}{1+c/n^+}}\right],
\ee
where  ${\cal T}_0 : = \nu_0 (\ln n^-/n^+)^{-1}$ is the apparent temperature in the absence of correlation, and we defined $n^-:=\sum_{i=1}^m \langle A_i(\nu_0) A_i^{\dag}(\nu_0)\rangle_{\rho_R} $, and $n^+:=\sum_{i=1}^m \langle A_i^{\dag}(\nu_0) A_i(\nu_0) \rangle_{\rho_R} $.

 The same comments made in the previous paragraph on the role of coherence are valid here as well. The achievable efficiency is increased if and only if 
\be\label{corrcond}
c (e^{\nu_0/{\cal T}_0}-1) \geq0,
\ee
 or in other words if and only if $c\geq 0$ when ${\cal T}_0 \geq 0$ (and the opposite when ${\cal T}_0 \leq 0$).

As illustrative example we mention $N$ two-level systems in Dicke states (Fig. \ref{refricorr}), 
 $|N,n_e\ket := \sqrt{\frac{n_e!}{N!n_g!}} \left(\sum_{i=1}^N \sigma_i^-\right)^{n_g} \otimes_{i=1}^N|e\ket_i $ \cite{dicke_coherence_1954,gross_superradiance:_1982}, where $n_e$ represents the number of delocalised excitations and $n_g:= N-n_e$ is the number of ground states. The corresponding maximal achievable efficiency is $\eta_{\rm ac} = \frac{T_C}{T_H-T_C} \left[1 -\frac{T_H}{{\cal T}_0} - \frac{T_H}{\nu_0} \ln{\frac{1+ n_e}{1+ n_g}}\right]$, where ${\cal T}_0:= \nu_0 \left(\ln{\frac{n_g}{n_e}}\right)^{-1}$. It appears then that the correlations present in the Dicke state increase the maximal achievable efficiency only for non-inverted states ($n_g\geq n_e$), equivalent to ${\cal T}_0 \geq 0$. As a comparison, we consider harmonic oscillators in a collective excitation state (analogue to Dicke states), and we obtain $\eta_{\rm ac} = \frac{T_C}{T_H-T_C} \left[1 -\frac{T_H}{{\cal T}_0} - \frac{T_H}{\nu_0} \ln{\frac{ n_e}{ n_e + N}}\right]$, where as previously $n_e$ stands for the number of collective excitations and $N$ is the number of harmonic oscillators. Interestingly, although the correlations increase with the number of excitations $n_e$, the impact on the apparent temperature shrinks away with $n_e$ (but the power is increased). \\

\begin{figure}[ht]
\centering
\includegraphics[width=4.5cm, height=4.5cm]{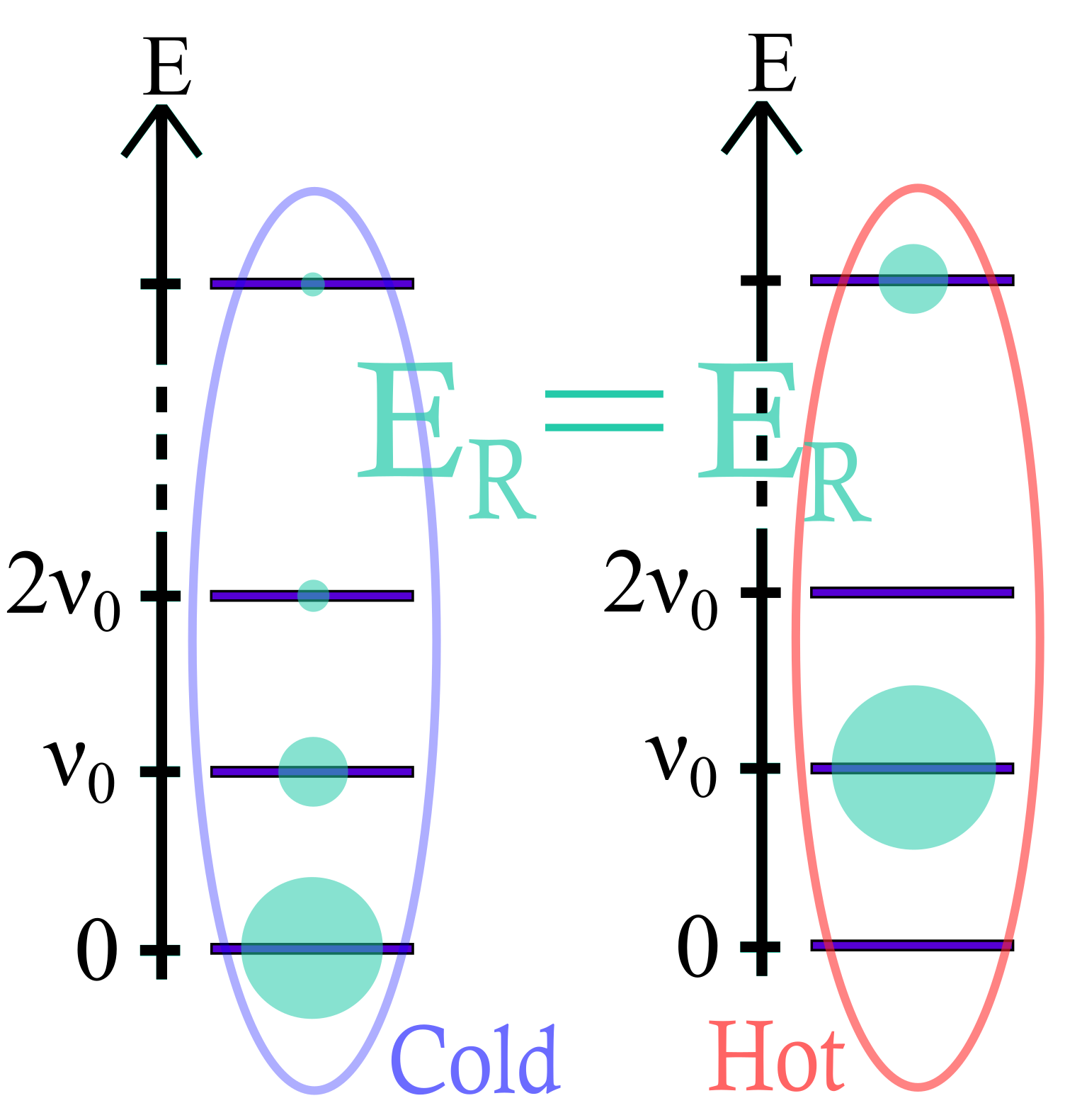}
\caption{Example of non-thermal population distribution (right-hand side) in a (N+1)-level system compared to a thermal distribution (left-hand side). Although both states have the same average energy $E_R$, the non-thermal distribution is ``hotter" than the thermal one.}
\label{nonth}
\end{figure}

{\bf Effect of non-thermal population distribution}.
No benefit from the above results can be obtained if $R$ is a non-degenerated single system. However, alternative non-thermal features can be found to increase the apparent temperature and therefore the maximal achievable efficiency of the machine. For thermal states, temperature and mean energy are in one-to-one correspondence. By contrast, as long as $H_R$ is not proportional to $ A_R^{\dag}(\nu_0)A_R(\nu_0)$, the apparent temperature is not determined by the average energy $E_R$. This implies that, differently from thermal states, non-thermal states of same energy $E_R$ can have {\it different apparent temperature}. 
Then, non-thermal population distributions can appear hotter (lower inverse apparent temperature) than thermal states of same energy $E_R$, providing therefore a higher efficiency (see Fig. \ref{nonth}). For a system of $N+1$ levels (of transition energy $\nu_0$), the apparent temperature can be re-written as ${\cal T}_R = \nu_0 \left(\ln{\frac{1-\rho_N}{1-\rho_0}}\right)^{-1}$, where $\rho_N$ and $\rho_0$ denote the populations in the most excited state and in the ground state, respectively, yielding a maximal achievable efficiency of 
\be
\eta_{\rm ac} = \frac{T_C}{T_H-T_C} \left[1 - \frac{T_H}{\nu_0} \ln{\frac{1- \rho_N}{1-\rho_0}}\right].
\ee

  As an example, for a 3-level system, one can exhibit non-thermal states with values of ${\cal T}_R/\nu_0$ up to $50 \%$ higher than the temperature of thermal states of same average energy $E_R$. For $N\geq 4$ this is even more drastic since one can find non-thermal states with ${\cal T}_R$ equal to infinity or even negative 
  while their thermal state counterparts of same energy have positive and finite temperatures (see Supplementary Information). 

We look now at the most common infinite-level system, the harmonic oscillator (of frequency $\nu_0$). The eigenoperators are the usual operator of creation and annihilation, $a = A_R(\nu_0)$ and $a^{\dag}=A_R^{\dag}(\nu_0)$. Then, $H_R$ is proportional to $A^{\dag}_R(\nu_0)A_R(\nu_0) = a^{\dag} a$, which implies that the apparent temperature and the maximal achievable efficiency are completely determined by $E_R$. In other words, there is no special effects from non-thermal states. This points out that no matter under which form the energy is stored in $R$, it powers the refrigerator with the same efficiency. In particular, the same increase of efficiency achieved by squeezing can be achieved with temperature increase by investing {\it the same amount of energy}: squeezed states are as efficient as thermal states.\\

{\bf Energy Extraction}. 
The above considerations can be straightforwardly extended to energy extraction regime. Such machines operate with opposite heat flows as compared with the above refrigerators, and thus the extracting conditions are found to be opposite to the refrigeration condition \eqref{refricond1},
\be\label{heatingcond}
\omega_0 \geq \nu_0 \frac{T_C}{T_H-T_C}\left(1-\frac{T_H}{{\cal T}_R}\right).
\ee 
 The efficiency $\eta_e$ of the energy extraction process is naturally defined by the amount of extracted energy, accounted by $\dot{E}_R$, divided by the energy invested, accounted by $\dot{Q}_{SR/H}$, $\eta_e := \dot{E}_R/\dot{Q}_{SR/H}$. From \eqref{mtsseq} one obtains the following expression for the efficiency, $\eta_e = \nu_0/(\omega_0+\nu_0) + {\cal O}(g^3/\nu_0^3)$, which, as for the refrigeration regime, is constant. The corresponding upper bound follows directly from \eqref{heatingcond}, 
 \be
 \eta_e \leq \left(1-\frac{T_C}{T_H}\right)\frac{{\cal T}_R}{{\cal T}_R -T_C},
 \ee
and is determined again by the apparent temperature of $R$ (assumed larger than $T_C$, otherwise the energy extraction is trivial). The above results and comments valid for refrigeration are also applicable in the energy extraction regime. 
Moreover, $R$ is expected to reach a steady state characterised by $\dot E_R =0$ which corresponds to an apparent temperature equal to $\nu_0[(\omega_0+\nu_0)/T_H-\omega_0/T_C]^{-1}$ (taking value outside of the interval $[T_C;T_H]$). Therefore, one can see that the higher the efficiency, the lower the steady state apparent temperature and the sooner the machine stops extracting energy (and conversely). This effect together with the interplay with bath-induced coherences and correlations in the steady state deserves further investigations.  

\subsection*{Discussion}

 We explore autonomous machines, a platform naturally adapted to investigate quantum effects in thermodynamic tasks. 
The existing designs of autonomous thermal machines are broadly extended (from quantum batteries made of single harmonic oscillators to arbitrary system of single energy transition or even discrete spectrum). In order to inspect more deeply perspectives of quantum boosts potentially hidden in previous upper bounds based on the Second Law we establish the actual efficiency of autonomous machines. This enables us to determine the maximal achievable efficiency. 
Thanks to the apparent temperature introduced in \cite{latune_apparent_2018}, it takes a simple an physically meaningful expression: the maximal achievable efficiency corresponds to the Carnot bound substituting the usual temperature by the apparent temperature of the quantum battery. Beyond increasing the validity and relevance of the apparent temperature, this happens to be an enlightening tool to understand the role of coherence, many-body correlations, and additional non-thermal characteristics in autonomous machines. Namely, as shown in \cite{latune_apparent_2018}, coherence and correlations contribute to heat flows together with the populations. Consequently, the usual population balance is modified which results in affecting the apparent temperature. The overall balance on the achievable efficiency is given in \eqref{effcoh} and \eqref{effcorr}, providing conditions for coherence and correlations to be beneficial, but also quantifying the resulting advantages. These results together with the broad validity of our model is expected to be fundamental in the realisation of autonomous thermal machines boosted by quantum and non-thermal effects.

\subsection*{Methods}
{\bf Upper bound from the Second Law}. We detail in this paragraph the performances that can be ideally expected from such machines according to the Second Law of Thermodynamics. 
Under the Markovian approximation and following \cite{alicki_quantum_1979, kosloff_quantum_2013, strasberg_quantum_2017} the Second Law applied to the ensemble $SR$ interacting with baths $C$ and $H$ takes the form of Spohn's inequality \cite{spohn_entropy_1978},
\be\label{markov2law}
\dot{S}(\rho_{SR}) -{\dot{Q}_{SR/C}\over T_C} -{\dot{Q}_{SR/H}\over T_H} \geq 0,
\ee 
where $S(\rho_{SR})=-{\rm Tr}\{\rho_{SR}\ln\rho_{SR}\}$ is the von Neumann entropy of the state $\rho_{SR}$, and $T_H$ and $T_C$ are the temperatures of $H$ and $C$, respectively.
For $S$ in a steady state (as required for a continuous machines \cite{levy_quantum_2012,kosloff_quantum_2013, kosloff_quantum_2014, gelbwaser-klimovsky_heat-machine_2014,gelbwaser-klimovsky_work_2015}), we derive from the energy conservation (First Law) and from the Second Law \eqref{markov2law} the following upper bound for the efficiency,
\be\label{2dlawupperbound}
\eta \leq {T_C\over T_H-T_C}\left(1 + T_H{\dot{S}(\rho_{R})\over -\dot{E}_R }\right),
\ee   
where $\rho_R:={\rm Tr}_S\rho_{SR}$.

Interestingly, this expression shows explicitly the decoupling of entropy and energy for non-thermal states. The two usual situations can also be recovered from \eqref{2dlawupperbound}. Namely, when $R$ provides energy without any entropy cost, corresponding to $\dot{S}(\rho_{R}) = 0$ and usually identified as {\it work} (Fig. \ref{thermalmachines}, panel a), and when $R$ is a thermal bath, where energy and entropy are {\it bound} by the relation $\dot{S}(\rho_{R}) = \dot{E}_R/T_R$, which corresponds to the absorption refrigerator (Fig. \ref{thermalmachines}, panel b). In both situations the corresponding Carnot bound is recovered from \eqref{2dlawupperbound}. The present study goes beyond these two usual situations (Fig.  \ref{thermalmachines}, panel c) and shows how quantum coherence, many-body correlations, and non-thermal population distribution can be exploited to boost autonomous thermal machines.
The upper bound \eqref{2dlawupperbound} was already derived in \cite{boukobza_breaking_2013, gelbwaser-klimovsky_heat-machine_2014,gelbwaser-klimovsky_work_2015}. However, its attainability is not proven, and
as discussed in the main text, there are several reasons to believe that it is not achievable.\\

{\bf Why dispersive coupling?} We consider the action of the baths $H$ and $C$ denoted collectively by $B$ on the ensemble $SR$ and adapt the formalism of Breuer and Petruccione's book \cite{breuer_theory_2007} to derive corrections to the local approach of the master equation of the reduced dynamic of $\rho_{SR}$ \cite{trushechkin_perturbative_2016}. Assuming the bath couplings are weak ($\lambda_C$ and $\lambda_H$ much smaller than the inverse of the bath correlation times), we start from the Redfield equation incremented with Markov approximation. This corresponds to Eq. (3.118) in \cite{breuer_theory_2007}, (setting $\hbar =1$), 
\be\label{markovredfield}
{d\over dt}\rho_{SR}^I(t) = - \int_0^{\infty} ds {\rm Tr}_B [V_{SB}^I(t),[V_{SB}^I(t-s),\rho_{SR}^I(t)\otimes\rho_B]],
\ee
 where the superscript $I$ denotes the interaction picture with respect to $H_{SR} + H_B$, namely $\rho_{SR}^I(t) := e^{iH_{SR}t}\rho_{SR}e^{-iH_{SR}t}$, $V_{SB}^I(t):= e^{i(H_{SR}+H_B)t}V_{SB}e^{-i(H_{SR}+H_B)t} = A_S^I(t)\tilde{A}_B(t)$, with $A_S^I(t) := e^{iH_{SR}t} A_S e^{-iH_{SR}t}$. We introduced the different notation of operators bearing tilde to denote interaction picture {\it with respect to the free Hamiltonian} of the corresponding subsystems, so that $\tilde{A}_B(t) := e^{iH_B t}A_B e^{-iH_B t}$. In the following we consider that $\lambda_H$ ($\lambda_C$) is included in $A_H$ ($A_C$). We define the rate of change of $\rho_{SR}^I$ only due to the interaction with the bath $j$ as
\be\label{jcontribution}
\dot\rho_{SR|_j}^I(t) := - \! \int_0^{\infty} \!\!\!\!ds {\rm Tr}_j [V_{Sj}^I(t),[V_{Sj}^I(t-s),\rho_{SR}^I(t)\otimes\rho_j]],
\ee
where $\rho_j$ denotes the density matrix of the bath $j$ and $V^I_{Sj}(t-s)$ denotes the operator $V_{Sj}:= A_SA_j$ (with $\lambda_j$ included within $A_j$) in the interaction picture defined above. One should note that since the baths are assumed to be thermal (and to remain independent), the following equality holds,
\be
\dot\rho_{SR}^I(t)= \dot\rho_{SR|_C}^I(t)+ \dot\rho_{SR|_H}^I(t).
\ee

We are ultimately interested in the heat flow from the bath  $j=H,C$ entering $SR$ defined in \eqref{defhflow}. Injecting the above expression \eqref{jcontribution} of $\dot \rho_{SR|_j}^I$ in \eqref{defhflow} one obtains,
\bea
\dot{Q}_{SR/j}&:=& - \int_0^{\infty} ds  \bra \tilde A_j(s)A_j \ket_{\rho_j} {\rm Tr}_{SR} \rho_{SR}^I(t)[H_{SR},A_S^I(t)]A_S^I(t-s) + c.c.
\eea
In the reminder of the Supplementary Material (as well as in the main text), we use the notation $\bra {\cal O}\ket_\rho $ to denote the expectation value of the operator ${\cal O}$ in the state $\rho$. Interestingly, the commutator appearing in the above expression can be calculated as follow,
\bea
[H_{SR},A_S^I(t)] &=& e^{iH_{SR}t}[H_{SR},A_S]e^{-iH_{SR}t}= e^{iH_{SR}t}([H_S,A_S] + gA_R [N_S,A_S])e^{-iH_{SR}t}.\nn\\
\eea
Then, it appears that if $N_S$ (the observable realising the coupling between $S$ and $R$) commutes with $A_S$ (the observable involved in the coupling between $S$ and the baths), the heat flows between $SR$ and the baths do not involve actively $R$. In other words one would have $\dot{Q}_{SR/j}=\dot{Q}_{S/j}$, meaning that the energy contained in $R$ and the correlations $V_{SR}$ is constant, which does not seem very promising for a thermal machine powered by $R$. 
 More generally, if $[N_S,A_S] = \kappa{\mathbb I}$, where ${\mathbb I}$ stands for the identity and $\kappa$ is a complex number, the heat flow becomes 
 \bea
\dot{Q}_{SR/j}&=& \dot Q_{S/j}- g \kappa \int_0^{\infty} ds  \bra \tilde A_j(s)A_j \ket_{\rho_j} {\rm Tr}_{SR} \rho_{SR}^I(t)A_R^I(t)A_S^I(t-s) + c.c.
\eea
 In the second term of the right-hand side, contributions of first and second order in $g$ are rapidly oscillating (at combinations of frequencies $\omega_0$ and $\nu$) so that they cancel out on average (secular approximation). Consequently, we obtain $\dot Q_{SR/j} = \dot Q_{S/j} + {\cal O}(g^3/|\nu|^3)$, which again is problematic for a thermal machine.     
  For each kind of system $S$ (in particular for $S$ a two-level system or a harmonic oscillator), there are several possible choices for $N_S$ to avoid $[N_S,A_S] = \kappa{\mathbb I}$. However, the only choice that avoid the problem simultaneously for both harmonic oscillators and two-level systems (and therefore for any other systems) is $N_S = \alpha H_S$ where $\alpha$ is a real number. For this reason we choose such coupling (dispersive coupling) between $S$ and $R$.\\

{\bf Expression of the bath dissipative operators}. The derivation of the reduced dynamics of $SR$ (following \cite{breuer_theory_2007}) requires the decomposition of $A_S^I(t)$ in the form $A_S^I(t) = \sum_{\Omega} e^{-i\Omega t} {\cal A}(\Omega)$ where the operators ${\cal A}(\Omega)$ are eigenoperators of $H_{SR}$ (also called ladder operators) verifying the relation $[H_{SR},{\cal A}(\Omega)] = -\Omega {\cal A}(\Omega)$, and the sum runs over an ensemble of frequencies $\Omega$ generated by combinations of the transition frequencies $\omega$ and $\nu$ of $S$ and $R$, respectively. The difficulty here comes from the interaction term $V_{SR}$ which turns the Hamiltonian $H_{SR}$ non-diagonalisable. As a consequence such eigenoperator decomposition is only approximately accessible through an expansion in term of the coupling strength $g$ between $S$ and $R$. We detail in the following the main steps leading to such decomposition. We start with 
\bea
A_S^I(t) &=& e^{i H_{SR}t} A_S e^{-i H_{SR}t}\nn\\
&=& \exp{\left\{ig\bar{\cal T} \int_0^t du N_S\tilde{A}_{R}(u)\right\}}e^{iH_St}A_S  e^{-iH_S t}\exp{\left\{-ig{\cal T} \int_0^t du N_S\tilde{A}_{R}(u)\right\}},
\eea
where ${\cal T}$ and $\bar{\cal T}$ denote the time ordering and anti-chronological ordering operators respectively (and $\tilde{A}_R(u) := e^{iH_R u} A_R e^{-iH_R u}$). For the system $X=S, R$, we decompose $A_X$ in a sum of eigenoperators $A_X(\nu)$ \cite{breuer_theory_2007} as introduced after \eqref{mtdotqc}, so that $\tilde{A}_X(u) = \sum_{\nu \in {\cal E}_X} A_X(\nu) e^{-i\nu u}$, the sum running over the authorised transition frequencies of $X$, denoted by the ensemble ${\cal E}_X$.
The eigenoperators verify $[H_X,A_X(\nu)] = -\nu A_X(\nu)$ and $A_R(-\nu) = A_R^{\dag}(\nu)$ (Hermicity of $A_R$) which implies that if $\nu$ belongs to ${\cal E}_R$ then, $-\nu$ too.  

 To obtain the final expression of the heat flow \eqref{mtdotqc} we restrict $S$ to be a two-level system or a harmonic oscillator but for now we continue with the general case as simplify the notations. 
  Then, assuming $g/|\nu| \ll 1$ for all $\nu \in {\cal E}_R$ and retaining only contributions up to second order in $g/|\nu|$, $A_S^I(t)$ can be rewritten as
\bea\label{intermedmsi}
A_S^I(t) &=& \sum_{\omega \in {\cal E}_S} e^{-i\omega t} \exp{\left\{ig\bar{\cal T} \int_0^t du N_S\tilde{A}_{R}(u)\right\}}A_S(\omega)\exp{\left\{-ig{\cal T} \int_0^t du N_S\tilde{A}_{R}(u)\right\}} \nn\\
&=& \sum_{\omega \in {\cal E}_S} e^{-i\omega t} A_S(\omega)\Bigg\{   1 +g\alpha\omega\sum_{\nu \in {\cal E}_R}A_R(\nu)\frac{e^{-i\nu t}-1}{\nu}\nn\\
&& -g^2\sum_{\nu_1 \ne -\nu_2 \in {\cal E}_R}\left[\alpha^2\omega^2A_R(\nu_2)A_R(\nu_1)-\alpha\omega N_S[A_R(\nu_2),A_R(\nu_1)]\right]\left[\frac{e^{-i\nu_1 t}}{\nu_1\nu_2} -\frac{e^{-i(\nu_1+\nu_2)t}}{\nu_2(\nu_1+\nu_2)} -\frac{1}{\nu_1(\nu_1+\nu_2)}\right]\nn\\
&& -g^2\sum_{\nu \in {\cal E}_R}\left[\alpha^2\omega^2A_R^{\dag}(\nu)A_R(\nu)-\alpha\omega N_S[A_R^{\dag}(\nu),A_R(\nu)]\right]\left[\frac{1-i\nu t -e^{-i\nu t}}{\nu^2}\right] + {\cal O}(g^3/|\nu|^3)\Bigg\}.
\eea
The operator $N_S$ being proportional to $H_S$, the commutator of $N_S$ with $A_S(\omega)$ is $[N_S,A_S(\omega)] = -\alpha \omega A_S(\omega)$. Rearranging \eqref{intermedmsi} we finally have a form resembling the eigenoperator decomposition we are looking for,
\be\label{spectraldecomp}
A_S^I(t) =\sum_{\omega \in {\cal E}_S} \left[{\cal A}(\omega) + {\cal C}(\omega) t \right] e^{-i\omega t} + \sum_{\omega \in {\cal E}_S, \nu \in {\cal E}_R} {\cal A}(\omega +\nu)e^{-i(\omega+\nu)t} + {\cal O}\left(\frac{g^2}{\nu^2}\right),
\ee
where for $\omega \in {\cal E}_S$,
\bea
{\cal A}(\omega)& =& A_S(\omega)\Bigg[1-\sum_{\nu \in {\cal E}_R}\frac{g\alpha\omega}{\nu}A_R(\nu) +  g^2\sum_{\nu_1 \ne -\nu_2 \in {\cal E}_R}\frac{1}{\nu_1(\nu_1+\nu_2)}\left[\alpha^2\omega^2A_R(\nu_2)A_R(\nu_1)-\alpha\omega N_S[A_R(\nu_2),A_R(\nu_1)]\right]\nn\\
&& \hspace{0.5in} -g^2\sum_{\nu \in {\cal E}_R}\frac{\alpha^2\omega^2}{\nu^2}A_R^{\dag}(\nu)A_R(\nu) + {\cal O}\left(\frac{g^3}{|\nu|^3}\right)\Bigg], \label{aw}\\
 {\cal C}(\omega) &=& ig^2\alpha^2 [H_S^2,A_S(\omega)]\sum_{\nu \in {\cal E}_R}\frac{1}{\nu}[A_R^{\dag}(\nu),A_R(\nu)],\label{coefc}
\eea
and for $\omega \in {\cal E}_S$ and $\nu \in {\cal E}_R$,
\be\label{awv}
{\cal A}(\omega+\nu) = \frac{g\alpha\omega}{\nu} A_S(\omega)A_R(\nu) +{\cal O}\left(\frac{g^2}{\nu^2}\right).
\ee

One can verify the following identities, ${\cal A}(-\omega)= {\cal A}^{\dag}(\omega)$, ${\cal C}(-\omega)= {\cal C}^{\dag}(\omega)$, and ${\cal A}(-\omega-\nu)= {\cal A}^{\dag}(\omega+\nu)$. Terms of second order in $g/|\nu|$ have already been dropped in Eq. \eqref{awv} and Eq. \eqref{spectraldecomp} since they generate in the master equation only terms of order 3 and 4.
The injection of the decomposition \eqref{spectraldecomp} into Eq. \eqref{markovredfield} brings the following master equation which takes into account the $SR$ coupling up to second order processes,
\bea\label{supmesr}
\dot{\rho}_{SR}^I &=& \sum_{\omega \in {\cal E}_S} \Gamma(\omega)\left[ {\cal A}(\omega)\rho_{SR}^I{\cal A}^{\dag}(\omega) - {\cal A}^{\dag}(\omega){\cal A}(\omega) \rho_{SR}^I \right] +{\rm h.c.}\nn\\ 
&&+ \sum_{\omega \in {\cal E}_S, \nu \in {\cal E}_R} \Gamma(\omega+\nu)\left[ {\cal A}(\omega+\nu)\rho_{SR}^I{\cal A}^{\dag}(\omega+\nu) - {\cal A}^{\dag}(\omega+\nu){\cal A}(\omega+\nu) \rho_{SR}^I \right] + {\rm h.c}\nn\\
&& + \Lambda_t(\rho_{SR}^I),
\eea
 where the secular approximation \cite{breuer_theory_2007} was performed for the terms involving operators ${\cal A}_S$, leading to the first two lines of \eqref{supmesr}. However, for terms involving ${\cal C}(\omega) t$ the secular approximation is not valid due to the time dependence. 
 The resulting map is denoted by $\Lambda_t$ and is given by
  \bea\label{map}
  \Lambda_t (\rho_{SR}^I) &=& - i \sum_{\omega \in {\cal E}_S} \partial_{\omega} \Gamma(\omega) [A_S^{\dag}(\omega) {\cal C}(\omega) \rho_{SR}^I -{\cal C}(\omega)\rho_{SR}^IA_S^{\dag}(\omega)] + {\rm h.c.} \nn\\
  &&+ \sum_{\omega,\omega' \in {\cal E}_S} \Gamma(\omega)e^{i(\omega' -\omega)t}  t [{\cal C}(\omega) \rho_{SR}^I A_S^{\dag}(\omega') -A_S^{\dag}(\omega'){\cal C}(\omega)\rho_{SR}^I + A_S(\omega)\rho_{SR}^I{\cal C}^{\dag}(\omega') -{\cal C}^{\dag}(\omega')A_S(\omega)\rho_{SR}^I]+{\rm h.c.}\nn\\
 && +{\cal O}\left({g^3\over |\nu|^3}\right),
  \eea
 where the complex function $\Gamma(\omega)$ is defined by $\Gamma(\omega) := \Gamma_C(\omega)+\Gamma_H(\omega)$ and  
 \be\label{defgamma}
 \Gamma_j (\omega) := \int_0^{\infty} ds e^{i\omega s}{\rm Tr}_j\rho_j A^I_j(s) A^I_j(0),
 \ee
 is of second order in the coupling coefficient $\lambda_j$, for $j = C,H$. We also define the spectral density of the bath $j$, 
\be\label{defspectraldensity}
G_j(\omega) := \Gamma_j(\omega) + \Gamma^{*}_j(\omega),
\ee 
 and, $G(\omega):= G_H(\omega) + G_C(\omega)$. We mention the following useful identity valid for thermal baths ($k_{B}=1$), 
\be\label{identitythermalbath}
G_j(-\omega) = e^{-\omega/T_j}G_j(\omega).
\ee
 
  One should note the presence of the factor $ t$ in the second line of \eqref{map}. Its linear growth in time can end up dominating the low-order terms. We show in the following that its contribution to the energy flow cancels out. However, one could wonder what happen if similar factors appear in higher order terms. Firstly, similar cancellation may happen (however showing it systematically is a challenging task).
Secondly, one can easily show that terms of third order grow at most as $\frac{g^3}{|\nu|^3}|\nu| t$. Assuming the coupling between $S$ and $R$ smaller than the coupling with the baths, $g\ll G_C(\omega_0)$, such third order terms might become significant only for times much larger than $\tau_R := \left (G_C(\omega_0) \frac{g^2}{|\nu|^2}\right )^{-1} $ which is the evolution timescale of $R$ and of the energy flows.
Similar reasoning can be repeated for higher orders.
 As a consequence, the contribution from higher orders might become relevant only after $R$ delivered (or stored) a significant amount of energy.  
Finally, on a more physical ground, one can observe that higher order terms in \eqref{intermedmsi} should be relevant if $S$ and $R$ were interacting in a close dynamics, without the presence of the baths. However, the decohering action of the baths limits the coherence of $S$ and therefore limits also the correlation time between $S$ and $R$. For weak coupling $g\ll\lambda_C,\lambda_H$, this dissipates high order processes between $S$ and $R$. 
Under these conditions, we limit ourselves to expansions of second order in $g/\nu_0$. 
Considering that all transition energies $\nu$ of $R$ are of same order of magnitude, we write ${\cal O}\left({g^3\over \nu_0^3}\right)$ instead of ${\cal O}\left({g^3\over |\nu|^3}\right)$ (valid also for Section ``Results'').

 From \eqref{supmesr} and the decomposition of $\Gamma(\omega)$ in contributions from each bath one can rewrite the above master equation as 
 \be\label{jcontrib}
 \dot{\rho}^I_{SR} = ({\cal L}_C + {\cal L}_H)\rho^I_{SR},
 \ee
 where ${\cal L}_j$ is the dissipative operator of the bath $j$, obtained from \eqref{supmesr} and \eqref{map} by substituting $\Gamma(\omega)$ by $\Gamma_j(\omega)$.\\
 

{\bf Heat flows}. We define the heat flow from the bath $j=C, H$ to the system $S$, $R$, or $SR$, denoted by $X$ in the following, as 
\be\label{methodsdefhf}
 \dot{Q}_{X/j}:= {\rm Tr}_{SR}{\cal L}_j \rho_{SR}^I H_{X}.
 \ee
When $X=SR$ we recover the definition \eqref{defhflow}. 
The heat flows are computed by injecting the expression of the bath dissipative operators in the above heat flow definition \eqref{methodsdefhf}. The calculation leading to \eqref{mtdotqc} is not complex but very cumbersome. The full detail can be found in the Supplementary Information. The main idea guiding the derivation is that $S$ is brought in a time interval of the order $\tau_{es}:=[G_C(\omega_0)]^{-1}$ to a thermal state at temperature $T_C$ (up to corrections of order $g/\nu_0$) due to its resonant and direct interaction with $C$. This timescale $\tau_{es}$ is much smaller than $\tau_R=[G_C(\omega_0)g^2/\nu_0^2]^{-1}$ the evolution timescale of $R$ and of the energy flows. Then, it is legitimate to approximate $\rho_S^I(t)$ by a thermal state at temperature $T_C$ in terms of second order in $g/\nu_0$.
However, the expectation value $\bra A_S(\omega_0)A_S^{\dag}(\omega_0)\ket_{\rho_{S}^I(t)}$ appears in the expansion of the expression of the heat flow \eqref{methodsdefhf} and is not of second order in $g/\nu_0$. Substituting $\rho_S^I(t)$ by the thermal state at the temperature $T_C$ is therefore not legitimate. We determine the expectation value $\bra A_S(\omega_0)A_S^{\dag}(\omega_0)\ket_{\rho_{S}^I(t)}$ up to second order in $g/\nu_0$ when $S$ is a harmonic oscillator or a two-level system (see Supplementary Information). 
When $S$ is a harmonic oscillator, the resulting expression for the heat flow from the cold bath $C$ is (for $t \gg \tau_{es}$)
 \be\label{methodhtflow}
 \dot{Q}_{SR/C} = \omega_0{g^2\alpha^2\omega_0^2\over \nu_0^2}{G_C(\omega_0)G_H(\omega_0 + \nu_0) \over G_C(\omega_0)-G_C(-\omega_0)}  \left( e^{-{\omega_0\over T_C}}\langle A_R^{\dag}(\nu_0) A_R(\nu_0)\rangle_{\rho_R^I} -e^{-{\omega_0+\nu_0 \over T_H}}\langle A_R(\nu_0) A_R^{\dag}(\nu_0)\rangle_{\rho_R^I}\right) + {\cal O}(g^3/\nu_0^3).
 \ee
 For a two-level system the expression is the same expect for the denominator of the pre-factor which is change to $G_C(\omega_0) + G_C(-\omega_0)$ instead of $G_C(\omega_0) - G_C(-\omega_0)$.
For both systems the following identity holds,  
\bea\label{methoder}
\dot{E}_R &=& {\nu_0 \over \omega_0 +\nu_0}\dot{Q}_{SR/H}+ {\cal O}\left({g^3\over \nu_0^3}\right) = - {\nu_0\over \omega_0}\dot{Q}_{SR/C} + {\cal O}\left({g^3\over \nu_0^3}\right).
\eea
Furthermore, the rate of variation of $E_S$ is $\dot{E}_S = {\cal O}(g^3/\nu_0^3)$ for $t\gg \tau_{es}$ for both systems.

\acknowledgements
This  work  is  based  upon  research  supported  by  the
South  African  Research  Chair  Initiative  of  the  Department  of  Science  and  Technology  and  National  Research Foundation. CLL acknowledges the support of the College of Agriculture, Engineering and Science of the UKZN.

\section*{Authors Contribution Statement}
C.L.L wrote the main manuscript text and I.S and F.P reviewed the manuscript. 

\section*{Additional Information}
{\bf Competing Interests}: The authors declare no competing interests.

\section*{Supplementary Information}

\appendix
\numberwithin{equation}{section}

 \section{General expression of the heat flows}\label{appendixheatflow}
\subsection{Expression of $\dot{Q}_{SR/j}$}
 The heat flow from the bath $j=C,H$ is defined in the main text as $\dot{Q}_{SR/j} : = {\rm Tr}_{SR} {\cal L}_j \rho_{SR}^I H_{SR}$. The expression announced in the main text is obtained by inserting the expression of the dissipative operator ${\cal L}_j$ in the above definition. One recurrent approximation (sometimes called the adiabatic approximation), is to consider that $S$ is in a thermal state at temperature $T_C$ denoted by $\rho_S^{eq}$ due to the resonant and continuous contact with the cold bath $C$. Corrective terms to this approximation are of order $g/|\nu|$ and are time-dependent. However, assuming that $g\ll \lambda_C$, $S$ remains approximately in a thermal state at all times, ensuring therefore that these corrective terms remains very small. As a consequence, when dealing with a term of second order in $g/|\nu|$ we approximate $\rho_S^I$ by $\rho_S^{eq}$. By contrast, when dealing with terms of lower order, such approximation is not valid since we want a final result taking into account up to second order in $g/\nu$. We mention the following identity used throughout this Section,
\bea\label{Arhoeq}
A_S(\omega) \rho^{eq}_S = e^{-\omega/T_{C}}  \rho^{eq}_S A_S(\omega),  
\eea
derived from 
\bea\label{thermalidentity}
e^{H_S/T_{C}}A_S(\omega)e^{-H_S/T_{C}} &=& \sum_{n=0}^{\infty} \frac{1}{n!}\frac{1}{T^n_{C}}Ad^n_{H_S} A_S(\omega) \nn\\
&=& \sum_{n=0}^{\infty} \frac{1}{n!}\frac{1}{T^n_{C}}(-\omega)^n A_S(\omega)\nn\\
&=& e^{-\omega/T_{C}}A_S(\omega),
 \eea
 where $Ad^n_{M}N := [M, Ad^{n-1}_M N]$ and $Ad^0_M N = N$. We recall that the ladder operators are such that $[H_S,A_S(\omega)] = -\omega A_S(\omega)$ so that $Ad^n_{H_S} A_S(\omega) = (-\omega)^n A_S(\omega)$.\\

 Starting with
 \bea
 \dot{Q}_{SR/j} &=& {\rm Tr} \{{\cal L}_j \rho^I_{SR} H_{SR} \}\nn\\
 &=& \sum_{\omega \in {\cal E}_S} \Gamma_j(\omega){\rm Tr} \{\rho_{SR}^I[{\cal A}^{\dag}(\omega),H_{SR}]{\cal A}(\omega) \} +c.c. \label{t1}\\ 
&&+ \sum_{\omega \in {\cal E}_S, \nu \in {\cal E}_R} \Gamma_j(\omega+\nu){\rm Tr} \{\rho_{SR}^I[{\cal A}^{\dag}(\omega+\nu),H_{SR}]{\cal A}(\omega+\nu)\}  + c.c \label{t2}\\
&& + {\rm Tr} \{\Lambda_{j,t}(\rho_{SR}^I)H_{SR}\},\label{t3}
 \eea
 we provide in the following the contribution from each of the three terms present in the above expression of the heat flow.

 \subsubsection{Contribution from the term \eqref{t1}} 
 The commutator $[{\cal A}^{\dag}(\omega),H_{SR}]$ gives
 \bea
 [{\cal A}^{\dag}(\omega),H_{SR}] = [{\cal A}^{\dag}(\omega),H_S] +[{\cal A}^{\dag}(\omega),H_R] +[{\cal A}^{\dag}(\omega),V_{SR}] 
 \eea
 
 Using the expression of ${\cal A}(\omega)$ given in Method we find
 \be
 [{\cal A}^{\dag}(\omega),H_S] = -\alpha \omega {\cal A}^{\dag},
 \ee
 \be
  [{\cal A}^{\dag}(\omega),H_R] = -g\alpha \omega A_S^{\dag}(\omega) A_R \left(1+\sum_{\nu \in {\cal E}_R}\frac{g\alpha\omega}{\nu}A_R(\nu)\right) + g A_S^{\dag}(\omega)N_S \sum_{\nu \in {\cal E}_R}\frac{g\alpha \omega}{\nu}[A_R(\nu),A_R] +{\cal O}\left({g^3\over \nu^3}\right),
 \ee
 and,
 \bea
 [{\cal A}^{\dag}(\omega),V_{SR}]  &=& g\alpha \omega A_S^{\dag}(\omega) A_R  + g^2A_S^{\dag}(\omega)\sum_{\nu \in {\cal E}_R}\frac{1}{\nu}\left(\alpha^2\omega^2 A_R A_R(\nu)+\alpha\omega N_S [A_R,A_R(\nu)] \right) \nn\\
 &&- g^2A_S^{\dag}(\omega)\sum_{\nu \in {\cal E}_R}\frac{1}{\nu}\left(\alpha^2\omega^2 A^{\dag}_R(\nu) A_R(\nu)+\alpha\omega N_S [A^{\dag}_R(\nu),A_R(\nu)]\right) .
 \eea
 
 Adding together the three contributions the commutator $[{\cal A}^{\dag}(\omega),H_{SR}]$ is reduced to
 \bea
 [{\cal A}^{\dag}(\omega),H_{SR}] &=& -\omega {\cal A}^{\dag}(\omega)- g^2A_S^{\dag}(\omega)\sum_{\nu \in {\cal E}_R}\frac{1}{\nu}\left(\alpha^2\omega^2 A^{\dag}_R(\nu) A_R(\nu)+\alpha\omega N_S [A^{\dag}_R(\nu),A_R(\nu)]\right) +{\cal O}\left({g^3\over |\nu|^3}\right)\nn\\
 &=&-\omega {\cal A}^{\dag}(\omega)- g^2\alpha \omega \sum_{\nu \in {\cal E}_R}\frac{1}{\nu}\left(N_SA_S^{\dag}(\omega) A^{\dag}_R(\nu) A_R(\nu)-A_S^{\dag}(\omega) N_S A_R(\nu)A^{\dag}_R(\nu)\right) +{\cal O}\left({g^3\over |\nu|^3}\right).
 \eea
 
 Substituting in \eqref{t1} we obtain,
 \bea
 \eqref{t1} &=& -\sum_{\omega \in {\cal E}_S} \omega G_j(\omega) \langle {\cal A}^{\dag}(\omega){\cal A}(\omega)\rangle_{\rho_{SR}^I}\nn\\
  && - g^2\sum_{\omega \in {\cal E}_S, \nu \in {\cal E}_R} \Gamma_j(\omega) \frac{\alpha \omega}{\nu}{\rm Tr}\left\{ \rho_{SR}^I \left( N_SA_S^{\dag}(\omega) A^{\dag}_R(\nu) A_R(\nu)-A_S^{\dag}(\omega) N_S A_R(\nu)A^{\dag}_R(\nu)\right){\cal A}(\omega)\right\} + c.c. \nn\\
  &&+{\cal O}\left({g^3\over |\nu|^3}\right)
 \eea
 
The second line of the above expression of the term \eqref{t1} can be simplified as follows
 \bea
 &-g^2&\sum_{\omega \in {\cal E}_S, \nu \in {\cal E}_R} \Gamma_j(\omega) \frac{\alpha \omega}{\nu}{\rm Tr}\left\{ \rho_{SR}^I \left( N_SA_S^{\dag}(\omega) A^{\dag}_R(\nu) A_R(\nu)-A_S^{\dag}(\omega) N_S A_R(\nu)A^{\dag}_R(\nu)\right){\cal A}(\omega)\right\} + c.c.\nn\\
 && = -g^2\sum_{\omega \in {\cal E}_S, \nu \in {\cal E}_R} \Gamma_j(\omega) \frac{\alpha \omega}{\nu}{\rm Tr}\left\{ \rho_{SR}^I \left( N_SA_S^{\dag}(\omega) A^{\dag}_R(\nu) A_R(\nu)-A_S^{\dag}(\omega) N_S A_R(\nu)A^{\dag}_R(\nu)\right)A_S(\omega)\right\} + c.c. +{\cal O}\left({g^3\over |\nu|^3}\right)\nn\\
 && = -g^2\sum_{\omega \in {\cal E}_S, \nu \in {\cal E}_R} \Gamma_j(\omega) \frac{\alpha \omega}{\nu}{\rm Tr}\left\{ \rho_{SR}^I \left( N_SA_S^{\dag}(\omega) A^{\dag}_R(\nu) A_R(\nu)+A_S^{\dag}(\omega) N_S A^{\dag}_R(\nu)A_R(\nu)\right)A_S(\omega)\right\} + c.c. +{\cal O}\left({g^3\over |\nu|^3}\right)\nn\\
 && = -g^2\sum_{\omega \in {\cal E}_S, \nu \in {\cal E}_R} \Gamma_j(\omega) \frac{\alpha \omega}{\nu}{\rm Tr}\left\{ \rho_{SR}^I [N_S,A_S^{\dag}(\omega)]_+ A_S(\omega)A^{\dag}_R(\nu) A_R(\nu)\right\} + c.c. +{\cal O}\left({g^3\over |\nu|^3}\right)\nn\\
  && = -g^2\sum_{\omega \in {\cal E}_S, \nu \in {\cal E}_R} G_j(\omega) \frac{\alpha \omega}{\nu}{\rm Tr}\left\{ \rho_{SR}^I [N_S,A_S^{\dag}(\omega)]_+ A_S(\omega)A^{\dag}_R(\nu) A_R(\nu)\right\} +{\cal O}\left({g^3\over |\nu|^3}\right),\nn\\
 \eea
 where $[X,Y]_+:=XY+YX$ denotes the anti-commutator of $X$ and $Y$. Note that for the bath $H$ the above contribution is trivially null (since $G_H(\omega)=0$). For the bath $C$, since the above term is already issued from second-order processes we can approximate $\rho_S^I$ by $\rho_S^{eq}$ (so that $[N_S,\rho_{SR}^I] = {\cal O}(g/|\nu|)$) and use \eqref{Arhoeq}, and we get
\be
\sum_{\omega \in {\cal E}_S, \nu \in {\cal E}_R} G_C(\omega) \frac{\alpha \omega}{\nu}{\rm Tr}\left\{ \rho_{SR}^I [N_S,A_S^{\dag}(\omega)]_+ A_S(\omega)A^{\dag}_R(\nu) A_R(\nu)\right\} = {\cal O}(g/|\nu|)
\ee
 so that the term \eqref{t1} is reduced to
 \be
 \eqref{t1} = -\sum_{\omega \in {\cal E}_S} \omega G_C(\omega) \langle {\cal A}^{\dag}(\omega){\cal A}(\omega)\rangle_{\rho_{SR}^I}+{\cal O}\left({g^3\over |\nu|^3}\right),
 \ee
 for the bath $C$, and $\eqref{t1} = 0$ for the bath $H$.\\
 
Using the expression of ${\cal A}(\omega)$ indicated in Methods, we derive the following expression for ${\cal A}^{\dag}(\omega){\cal A}(\omega)$,
\bea
{\cal A}^{\dag}(\omega){\cal A}(\omega) &=& A^{\dag}_S(\omega)A_S(\omega) - A^{\dag}_S(\omega)A_S(\omega)\sum_{\nu_1,\nu_2 \in {\cal E}_R}\frac{g^2\alpha^2\omega^2}{\nu_1\nu_2}A_R(\nu_1)A_R(\nu_2) -2\sum_{\nu \in {\cal E}_R} \frac{g^2\alpha^2\omega^2}{\nu^2}A^{\dag}_S(\omega)A_S(\omega)A^{\dag}_R(\nu)A_R(\nu) \nn\\
&&+g^2\sum_{\nu_1 \ne -\nu_2 }\frac{1}{\nu_2(\nu_1+\nu_2)}\left(\alpha^2\omega^2 A_S^{\dag}(\omega)A_S(\omega)A_R(\nu_1)A_R(\nu_2) + \alpha\omega A_S^{\dag}(\omega)N_SA_S(\omega)[A_R(\nu_1),A_R(\nu_2)]\right)  \nn\\
&&+g^2\sum_{\nu_1 \ne -\nu_2 }\frac{1}{\nu_2(\nu_1+\nu_2)}\left(\alpha^2\omega^2 A_S^{\dag}(\omega)A_S(\omega)A_R(\nu_1)A_R(\nu_2) - \alpha\omega A_S^{\dag}(\omega)A_S(\omega)N_S[A_R(\nu_1),A_R(\nu_2)]\right) \nn\\
&&+{\cal O}\left({g^3\over |\nu|^3}\right)\nn\\
&=&   A^{\dag}_S(\omega)A_S(\omega)\Bigg[1 - \sum_{\nu_1,\nu_2 \in {\cal E}_R}\frac{g^2\alpha^2\omega^2}{\nu_1\nu_2}A_R(\nu_1)A_R(\nu_2) -2\sum_{\nu \in {\cal E}_R} \frac{g^2\alpha^2\omega^2}{\nu^2}A^{\dag}_R(\nu)A_R(\nu) \nn\\
&&+\sum_{\nu_1 \ne -\nu_2 }\frac{g^2\alpha^2\omega^2}{\nu_2(\nu_1+\nu_2)} [A_R(\nu_1),A_R(\nu_2)]_+ \Bigg]+{\cal O}\left({g^3\over |\nu|^3}\right)\nn\\
&=&   A^{\dag}_S(\omega)A_S(\omega)\Bigg[ 1 - \sum_{\nu_1,\nu_2 \in {\cal E}_R}\frac{g^2\alpha^2\omega^2}{\nu_1\nu_2}A_R(\nu_1)A_R(\nu_2) -2\sum_{\nu \in {\cal E}_R} \frac{g^2\alpha^2\omega^2}{\nu^2}A^{\dag}_R(\nu)A_R(\nu) \nn\\
&&+\sum_{\nu_1 \ne -\nu_2 }\frac{g^2\alpha^2\omega^2}{\nu_1\nu_2} A_R(\nu_1)A_R(\nu_2)\Bigg]+{\cal O}\left({g^3\over |\nu|^3}\right) \nn\\
&=&   A^{\dag}_S(\omega)A_S(\omega)\left[ 1 - \sum_{\nu \in {\cal E}_R} \frac{g^2\alpha^2\omega^2}{\nu^2}A^{\dag}_R(\nu)A_R(\nu)\right] +{\cal O}\left({g^3\over |\nu|^3}\right)
\eea

 Finally the  term \eqref{t1} is reduced to
  \bea
 \eqref{t1} &=& -\sum_{\omega \in {\cal E}_S}\omega G_j(\omega)\left[ \langle A_S^{\dag}(\omega)A_S(\omega)\rangle_{\rho_{SR}^I} -g^2\alpha^2\omega^2 \sum_{\nu \in {\cal E}_R} \frac{1}{\nu^2}  \langle A_S^{\dag}(\omega)A_S(\omega) A_R^{\dag}(\nu)A_R(\nu)\rangle_{\rho_{SR}^I}\right] +{\cal O}\left({g^3\over |\nu|^3}\right) \nn\\
  &=& -\sum_{\omega \in {\cal E}_S} \omega G_j(\omega) \langle A_S^{\dag}(\omega)A_S(\omega)\rangle_{\rho_{SR}^I}+{\cal O}\left({g^3\over |\nu|^3}\right)
    \eea
using the identity \eqref{Arhoeq} in the last line (since the terms are already of second order in $g/\nu$).

 \subsubsection{Contribution from the term \eqref{t2}}
 The commutator $[{\cal A}^{\dag}(\omega +\nu),H_{SR}] $ gives
 \be
 [{\cal A}^{\dag}(\omega +\nu),H_{SR}] = -(\omega+\nu){\cal A}^{\dag}(\omega +\nu),
 \ee
 so that 
 \bea
 \eqref{t2} &=& - \sum_{\omega \in {\cal E}_S, \nu \in {\cal E}_R} (\omega + \nu) \Gamma_j(\omega+\nu){\rm Tr} \{\rho_{SR}^I{\cal A}^{\dag}(\omega+\nu){\cal A}(\omega+\nu)\}  + c.c \nn\\
 &=& - \sum_{\omega \in {\cal E}_S, \nu \in {\cal E}_R} (\omega + \nu) G_j(\omega+\nu)\frac{g^{2}\alpha^2\omega^{2}}{\nu^2} \langle A^{\dag}_S(\omega)A_S(\omega)A_R^{\dag}(\nu)A_R(\nu)\rangle_{\rho_{SR}^I}    
 \eea

 \subsubsection{Contribution from the term \eqref{t3}}
 Using the expression of the dissipative map $\Lambda_{j,t}$ given in Methods, we have,
 \bea
\eqref{t3} &=& {\rm Tr}\{\Lambda_{j,t} (\rho_{SR}^I)H_{SR}\}\nn\\
 &=& -i \sum_{\omega \in {\cal E}_S} \partial_{\omega}\Gamma_j(\omega){\rm Tr}\{\rho_{SR}^I[H_{SR},A_S^{\dag}(\omega)]{\cal C}(\omega)\} + c.c. \nn\\
 &&+\sum_{\omega,\omega' \in {\cal E}_S} \Gamma_j(\omega) t e^{i(\omega'-\omega)t}  {\rm Tr}\left\{\rho_{SR}^I\Big([A_S^{\dag}(\omega'),H_{SR}]{\cal C}(\omega) + [{\cal C}^{\dag}(\omega'),H_{SR}]A_S(\omega)\Big)\right\} + c.c..
 \eea 
We are again dealing with terms from second-order so that we can use the approximation $\rho_S^I = \rho_S^{eq} + {\cal O}(g/|\nu|)$ and the relations $[H_{SR},A_S^{\dag}(\omega)] = \omega A^{\dag}_S(\omega) + {\cal O}(g/|\nu|)$, $[A_S^{\dag}(\omega'),H_{SR}] = -\omega' A^{\dag}_S(\omega') + {\cal O}(g/|\nu|)$ and $ [{\cal C}^{\dag}(\omega'),H_{SR}] = -\omega' {\cal C}^{\dag}(\omega')+{\cal O}(g/|\nu|)$, yielding,
\bea
\eqref{t3}&=& -i \sum_{\omega \in {\cal E}_S} \omega \partial_{\omega}\Gamma_j(\omega){\rm Tr}\{\rho_{SR}^IA_S^{\dag}(\omega){\cal C}(\omega)\} +{\cal O}\left({g^3\over |\nu|^3}\right) + c.c. \nn\\
 &&-\sum_{\omega,\omega' \in {\cal E}_S} \omega' \Gamma_j(\omega) t e^{i(\omega'-\omega)t}  {\rm Tr}\left\{\rho_{SR}^I\Big(A_S^{\dag}(\omega'){\cal C}(\omega) + {\cal C}^{\dag}(\omega')A_S(\omega)\Big)\right\} +{\cal O}\left({g^3\over |\nu|^3}\right) +  c.c..
 \eea 
 We now derive a useful identity
 \bea
 {\rm Tr}\{\rho_{SR}^I{\cal C}^{\dag}(\omega')A_S(\omega)\} &=& -ig^2 \alpha^2 \sum_{\nu \in {\cal E}_R} \frac{1}{\nu} \langle [A_R(\nu)^{\dag},A_R(\nu)]\rangle_{\rho_R^I} {\rm Tr}\{\rho_S^I [A_S^{\dag}(\omega'),H_S^2]A_S(\omega)\}+{\cal O}\left({g^3\over |\nu|^3}\right)\nn\\
 &=& -ig^2 \alpha^2 \sum_{\nu \in {\cal E}_R} \frac{1}{\nu} \langle [A_R(\nu)^{\dag},A_R(\nu)]\rangle_{\rho_R^I} {\rm Tr}\{[H_S^2, A_S(\omega)\rho_S^I] A_S^{\dag}(\omega')\} +{\cal O}\left({g^3\over |\nu|^3}\right)\nn\\
 &=& -ig^2 \alpha^2 \sum_{\nu \in {\cal E}_R} \frac{1}{\nu} \langle [A_R(\nu)^{\dag},A_R(\nu)]\rangle_{\rho_R^I} {\rm Tr}\{[H_S^2, A_S(\omega)]\rho_S^IA_S^{\dag}(\omega')\} +{\cal O}\left({g^3\over |\nu|^3}\right)\nn\\
 &=& - {\rm Tr}\{\rho_{SR}^IA_S^{\dag}(\omega'){\cal C}(\omega)\} +{\cal O}\left({g^3\over |\nu|^3}\right),\label{CdagA}
 \eea
 where we again used $\rho_S^I = \rho_S^{eq} + {\cal O}(g/|\nu|)$. From the above identity \eqref{CdagA} one can simplify the term \eqref{t3} to
 \bea
 \eqref{t3} &=& -i \sum_{\omega \in {\cal E}_S}\omega \partial_{\omega}G_j(\omega)  {\rm Tr}\{\rho_{SR}^IA_S^{\dag}(\omega){\cal C}(\omega)\}+{\cal O}\left({g^3\over |\nu|^3}\right)\nn\\
 &=& -i \sum_{\omega \geq 0} \omega G'_j(\omega)  {\rm Tr}\{\rho_{SR}^IA_S^{\dag}(\omega){\cal C}(\omega)\} - \omega G'_j(-\omega)  {\rm Tr}\{\rho_{SR}^IA_S^{\dag}(-\omega){\cal C}(-\omega)\}+{\cal O}\left({g^3\over |\nu|^3}\right),
 \eea
 with the notation $G'_j(\omega):=\partial_{\omega} G_j(\omega) $.
 We recall that for the bath $H$ the contribution of the term \eqref{t3} is trivially null ($G_H(\omega)=0$ for all $\omega \in {\cal E}_S$).
For the bath $C$ the expression of \eqref{t3} can be simplified further. Using \eqref{Arhoeq} one can show in a similar way
 \be\label{ACdag}
 {\rm Tr}\{\rho_{SR}^IA_S(\omega){\cal C}^{\dag}(\omega)\} = -e^{\omega/T_C}  {\rm Tr}\{\rho_{SR}^IA_S^{\dag}(\omega){\cal C}(\omega)\}.
 \ee 
Furthermore, the equality
\be\label{partialGj}
 G'_j(-\omega) = e^{-\omega/T_j}[G_j(\omega)/T_j -G'_j(\omega)]
\ee
 holds for thermal baths. 
 Combining the two last equalities \eqref{ACdag} and \eqref{partialGj} we finally obtain,
 \bea
 \eqref{t3} &=& -i \sum_{\omega \geq 0} \omega \langle A_S^{\dag}(\omega){\cal C}(\omega)\rangle_{\rho_{SR}^I} [G'_C(\omega) + e^{\omega/T_C}G'_C(-\omega)]+{\cal O}\left({g^3\over |\nu|^3}\right)\nn\\
 &=& -i \sum_{\omega \geq 0}\frac{\omega}{T_C} \langle A_S^{\dag}(\omega){\cal C}(\omega)\rangle_{\rho_{SR}^I} G_C(\omega)+{\cal O}\left({g^3\over |\nu|^3}\right). \nn\\
 \eea
 From the expression of ${\cal C}(\omega)$ (in Methods) and using the identity \eqref{Arhoeq} one can rewrite \eqref{t3} in the form
 \be
 \eqref{t3}= -\sum_{\omega \in {\cal E}_S, \nu \in {\cal E}_R} 2 \nu T_j^{-1}\frac{\alpha^2\omega^2g^2}{\nu^2}G_j(\omega) \langle A_S^{\dag}(\omega) A_S(\omega)H_S A_R^{\dag}(\nu)A_R(\nu)\rangle_{\rho_{SR}^I}+{\cal O}\left({g^3\over |\nu|^3}\right),
 \ee 
 which also gives the right expression for $H$, namely $\eqref{t3}=0$.

 \subsubsection{Final expression for $\dot{Q}_{SR/j}$}
 Combining terms \eqref{t1}, \eqref{t2}, and \eqref{t3} we obtain finally, valid for $j=H,C$,
 \bea\label{gexprhfsr}
 \dot{Q}_{SR/j} &=& -\sum_{\omega \in {\cal E}_S} \omega G_j(\omega) \langle A_S^{\dag}(\omega)A_S(\omega)\rangle_{\rho_{S}^I}\nn\\
 && - \sum_{\omega \in {\cal E}_S, \nu \in {\cal E}_R} (\omega + \nu) G_j(\omega+\nu)\frac{g^{2}\alpha^2\omega^{2}}{\nu^2} \langle A^{\dag}_S(\omega)A_S(\omega)A_R^{\dag}(\nu)A_R(\nu)\rangle_{\rho_{SR}^I} \nn\\
 && -\sum_{\omega \in {\cal E}_S, \nu \in {\cal E}_R} 2 \nu T_j^{-1}\frac{\alpha^2\omega^2g^2}{\nu^2}G_j(\omega) \langle A_S^{\dag}(\omega) A_S(\omega)H_S A_R^{\dag}(\nu)A_R(\nu)\rangle_{\rho_{SR}^I} +{\cal O}\left({g^3\over|\nu|^3}\right).
 \eea
 Since we neglect contributions of order higher than $g^2/|\nu|^2$ one can once again approximate $\rho_S^I$ by $\rho_S^{eq}$ (or $\rho_{SR}^{eq}$ by $\rho_S^{eq}\rho_R^I$), yielding,
  \bea
  \langle A_S^{\dag}(\omega) A_S(\omega) A_R^{\dag}(\nu)A_R(\nu)\rangle_{\rho_{SR}^I} =  \langle A_S^{\dag}(\omega) A_S(\omega)\rangle_{\rho_{S}^{eq}} \langle A_R^{\dag}(\nu)A_R(\nu)\rangle_{\rho_{R}^I}   +{\cal O}\left({g\over|\nu|}\right),
 \eea
 and,
  \bea
  \langle A_S^{\dag}(\omega) A_S(\omega)H_S A_R^{\dag}(\nu)A_R(\nu)\rangle_{\rho_{SR}^I} =  \langle A_S^{\dag}(\omega) A_S(\omega)H_S\rangle_{\rho_{S}^{eq}} \langle A_R^{\dag}(\nu)A_R(\nu)\rangle_{\rho_{R}^I}   +{\cal O}\left({g\over|\nu|}\right).
 \eea
 However, the first line of \eqref{gexprhfsr} is of order 0 in $g/|\nu|$ so that we cannot approximate $ \langle A_S^{\dag}(\omega)A_S(\omega)\rangle_{\rho_{S}^I}$ by $ \langle A_S^{\dag}(\omega)A_S(\omega)\rangle_{\rho_{S}^{eq}}$. A derivation of an expression of $ \langle A_S^{\dag}(\omega)A_S(\omega)\rangle_{\rho_{S}^I}$ to second order in $g/|\nu|$ is presented in Section \ref{appendixeval}.\\
  
  \subsection{Expression of $\dot{Q}_{S/j}$}
 From the above calculations we can easily derive the following expression for $\dot{Q}_{S/j}$, the heat flow entering $S$ only, 
 \bea\label{gexprhfs}
 \dot{Q}_{S/j} &=& -\sum_{\omega \in {\cal E}_S} \omega G_j(\omega) \langle A_S^{\dag}(\omega)A_S(\omega)\rangle_{\rho_{S}^I}\nn\\
 && - \sum_{\omega \in {\cal E}_S, \nu \in {\cal E}_R} \omega  G_j(\omega+\nu)\frac{g^{2}\alpha^2\omega^{2}}{\nu^2} \langle A^{\dag}_S(\omega)A_S(\omega)\rangle_{\rho_{S}^{eq}} \langle A_R^{\dag}(\nu)A_R(\nu)\rangle_{\rho_{R}^I} \nn\\
 && -\sum_{\omega \in {\cal E}_S, \nu \in {\cal E}_R} 2 \nu T_j^{-1}\frac{\alpha^2\omega^2g^2}{\nu^2}G_j(\omega) \langle A_S^{\dag}(\omega) A_S(\omega)H_S \rangle_{\rho_{S}^{eq}} \langle A_R^{\dag}(\nu)A_R(\nu)\rangle_{\rho_{R}^I} +{\cal O}\left({g^3\over|\nu|^3}\right).
 \eea
The unique difference from the expression of $\dot{Q}_{SR/j}$ comes from the second line which carries an energy $\omega$ instead of $\omega + \nu$.

 \section{Evaluation of $\langle A_S^{\dag}(\omega)A_S(\omega)\rangle_{\rho_{S}^I}$ to ${\rm 2^d}$ order in $g/|\nu|$}\label{appendixeval}
 In Section \ref{appendixheatflow} we obtained expressions of the different heat flows up to order $g^2/\nu^2$. Accordingly, the expectation value  $\langle A_S^{\dag}(\omega)A_S(\omega)\rangle_{\rho_{S}^I}$ therein has to be evaluated to the same order $g^2/\nu^2$. To do so we take the time derivative of $\langle A_S^{\dag}(\omega)A_S(\omega)\rangle_{\rho_{S}^I}$ and keep only terms up to second order in $g/|\nu|$. The obtained differential equation can be easily solved when $S$ is a harmonic oscillator or a two-level system. In the remainder of the Supplemental Material we focus on such situations and denote by $\omega_0$ the transition frequency of $S$, implying ${\cal E}_S = \{-\omega_0,\omega_0\}$. To simplify further the model and the results we assume that $C$ is resonant only with $S$, and that $H$ is resonant with only one transition energy of $SR$, denoted by $\omega_0 +\nu_0$ with $\nu_0 \geq 0$. Then, $G_H(\omega_0 + \nu) = 0$ for all $\nu\ne\nu_0$. This corresponds to the model considered in the main text.
%
 Starting with
 \be
 \frac{d}{dt} \langle A_S^{\dag}(\omega)A_S(\omega)\rangle_{\rho_{S}^I} = {\rm Tr} \{ \dot{\rho}_S^I A_S^{\dag}(\omega)A_S(\omega)\}={\rm Tr} \{ \dot{\rho}_{SR}^I A_S^{\dag}(\omega)A_S(\omega)\},
 \ee
 we inject expression from Methods of $\dot{\rho}_{SR}^I$ in the above one and after similar calculations as in the last section \ref{appendixheatflow} we obtain
 \bea
 \frac{d}{dt} \langle A_S^{\dag}(\bar\omega)A_S(\bar\omega)\rangle_{\rho_{S}^I} &=& \sum_{\omega = \pm\omega_0} G_C(\omega) \langle [A_S^{\dag}(\omega), A_S^{\dag}(\bar\omega)A_S(\bar\omega)]A_S(\omega)\rangle_{\rho_{S}^I}\nn\\
 &&+ \sum_{\omega +\nu =\pm(\omega_0 + \nu_0)} G_H(\omega+\nu)\frac{g^2\alpha^2\omega^2}{\nu^2} \langle A_R^{\dag}(\nu)A_R(\nu)\rangle_{\rho_R^I} \langle [A_S^{\dag}(\omega),A_S^{\dag}(\bar\omega)A_S(\bar\omega)]A_S(\omega)\rangle_{\rho_S^I} \nn\\
 && + \sum_{\omega=\pm\omega_0, \nu \in {\cal E}_R} G_C(\omega) g^2\alpha^2\frac{\omega}{\nu} T_C^{-1} \langle [A_R^{\dag}(\nu) , A_R(\nu)]\rangle_{\rho_R^I} \langle [A_S^{\dag}(\omega),A_S^{\dag}(\bar\omega)A_S(\bar\omega)]A_S(\omega)H_S\rangle_{\rho_S^I}\nn\\
 &&+{\cal O}\left({g^3\over |\nu|^3}\right). \label{eqdiff}
 \eea
 To continue we need to compute the commutators appearing in the differential equation. We do so in the following Sections when $S$ is a harmonic oscillator (Section \ref{sectionho}) and a two-level system (Section \ref{sectiontls}).

 \subsection{Harmonic oscillators}\label{sectionho}
 We assume here that $S$ is a harmonic oscillator of frequency $\omega_0$ and $a$, $a^{\dag}$ are the annihilation and creation operators. 
 Harmonic oscillators usually couple to baths through quadratures operators. We therefore choose $P_S$ of the following form,
 \be
 P_S = c a^{\dag} + c^{*} a,
 \ee
where $c$ is a complex number. 
 As a consequence, $A_S(\omega_0) = c^{*}a$ and $A_S(-\omega_0) = c a^{\dag}$. Equation \eqref{eqdiff} is then simplified to
 \bea
  \frac{d}{dt} \langle a^{\dag} a\rangle_{\rho_{S}^I} &=& -\lambda \langle a^{\dag} a\rangle_{\rho_{S}^I} + r,
 \eea
where
\be
\lambda = |c|^4 \left\{ G_C(\omega_0) - G_C(-\omega_0) +  \frac{g^2\alpha^2\omega_0^2}{\nu_0^2} \left[G_H(\omega_0+\nu_0) \langle A_R^{\dag}(\nu_0)A_R(\nu_0) \rangle_{\rho_R^I} - G_H(-\omega_0 - \nu_0)\langle A_R(\nu_0)A_R^{\dag}(\nu_0) \rangle_{\rho_R^I}\right] \right\},
 \ee
 and 
 \bea
 r &=& |c|^4 G_C(-\omega_0) + |c|^4 g^2\alpha^2\omega_0^2  \Bigg\{\frac{G_H(-\omega_0 -\nu_0)}{\nu_0^2} \langle A_R(\nu_0)A_R^{\dag}(\nu_0) \rangle_{\rho_R^I} -\left[G_C(\omega_0)\langle (a^{\dag} a)^2\rangle_{\rho_S^I} + G_C(-\omega_0)\langle aa^{\dag 2}a \rangle_{\rho_S^I}\right]
\nn\\
&&\hspace{9cm}\times\sum_{\nu \in {\cal E}_R} {2T_C^{-1}\over \nu}   \langle A_R^{\dag}(\nu)A_R(\nu) \rangle_{\rho_R^I}\Bigg\} + {\cal O}(g^3/|\nu|^3).
 \eea

 In the following we set $c=1$ since the phase of $c$ has no observable influence (and the amplitude of $c$ can be included in $G(\omega)$).  \\
 
 
 A quick analysis of the dynamics of $\langle A_R(\nu)A_R^{\dag}(\nu) \rangle_{\rho_R^I}$ reveals that its time derivative is of order 2 in $g/|\nu|$ (the contribution of ${\cal L}_0$ clearly gives zero whereas the contribution of $g{\cal L}_1$ turns out to be of order $g^2/\nu^2$ when $t\gg \tau_1:= G_C^{-1}(\omega_0)$. One obtains from this observation that $\lambda$ varies very slowly, at a rate of order $g^4/\nu^4$, so that it can safely be taken as constant. The same conclusion can be drawn for $r$. Then we conclude that a good approximation of $\langle a^{\dag} a\rangle_{\rho_{S}^I}$ is 
 \be
 \langle a^{\dag} a\rangle_{\rho_{S}^I} = e^{-\lambda t} \langle a^{\dag} a\rangle_{\rho_{S}(0)} + { 1-e^{-\lambda t} \over \lambda} r =_{t\gg \tau_1} {r \over \lambda}.
 \ee  
  
 Substituting in \eqref{gexprhfsr}, \eqref{gexprhfs}, \eqref{gexprhfr} and using \eqref{Arhoeq} one obtains for the heat flows
 \be
 \dot{Q}_{SR/C} = \omega_0{g^2\alpha^2\omega_0^2\over \nu_0^2}{G_C(\omega_0)G_H(\omega_0 + \nu_0) \over G_C(\omega_0)-G_C(-\omega_0)}  \left( e^{-{\omega_0\over T_C}}\langle A_R^{\dag}(\nu_0) A_R(\nu_0)\rangle_{\rho_R^I} -e^{-{\omega_0+\nu_0 \over T_H}}\langle A_R(\nu_0) A_R^{\dag}(\nu_0)\rangle_{\rho_R^I}\right) + {\cal O}(g^3/\nu_0^3),
 \ee
 \bea
 \dot{Q}_{SR/H}  &=&  -(\omega_0+\nu_0){g^2\alpha^2\omega_0^2\over \nu_0^2}{G_C(\omega_0)G_H(\omega_0 + \nu_0) \over G_C(\omega_0)-G_C(-\omega_0)}  \left( e^{-{\omega_0\over T_C}}\langle A_R^{\dag}(\nu_0) A_R(\nu_0)\rangle_{\rho_R^I} -e^{-{\omega_0+\nu_0 \over T_H}}\langle A_R(\nu_0) A_R^{\dag}(\nu_0)\rangle_{\rho_R^I}\right) + {\cal O}(g^3/\nu_0^3)\nn\\
 &=&- {\omega_0 +\nu_0 \over \omega_0} \dot{Q}_{SR/C} + {\cal O}\left({g^3\over \nu_0^3}\right), \label{horelationhfxy}
 \eea
 \be\label{hoqsx}
 \dot{Q}_{S/C} = -\dot{Q}_{S/H}+ {\cal O}\left({g^3\over \nu^3}\right) = \dot{Q}_{SR/C}+ {\cal O}\left({g^3\over \nu^3}\right).
 \ee
Equation \eqref{hoqsx} implies in particular that $\dot{E}_S =  {\cal O}\left({g^3\over \nu_0^3}\right)$. This is the usual condition of steady state for continuous thermal machines \cite{levy_quantum_2012,kosloff_quantum_2013, kosloff_quantum_2014, gelbwaser-klimovsky_heat-machine_2014} mentioned in the main text and valid for times $t$ much bigger than $\tau_{es}= [G_C(\omega_0)]^{-1}$. \\

 \subsection{Two-level systems}\label{sectiontls}
 We assume is this section that $S$ is a two-level system of transition frequency $\omega_0$. Such system usually couples to baths through the operators $\sigma_+$ and $\sigma_-$ (the Pauli matrices) so that a natural choice for $P_S$ is 
 \be
 P_S = c\sigma_+ + c^{*} \sigma_-.
 \ee
 where $c$ is a complex number. As a consequence $A_S(\omega_0) = c^{*} \sigma_-$, $A_S(-\omega_0) = c \sigma_+$, $\langle A_S(\omega_0) A_S^{\dag}(\omega_0) \rangle_{\rho_S^I} = |c|^2 \rho_{gg}$, and $\langle A_S^{\dag}(\omega_0) A_S(\omega_0) \rangle_{\rho_S^I} = |c|^2 \rho_{ee}$, where $\rho_{gg} := \langle g|\rho_S^I |g\rangle$ and $\rho_{ee} := \langle e|\rho_S^I |e\rangle$, being $|g\rangle$ and $|e\rangle $ the ground and excited state of $S$ respectively. \\
 From \eqref{eqdiff} we obtain the following dynamics
 \be
 \dot{\rho}_{ee} = -\dot{\rho}_{gg} = - R_+ \rho_{ee} + R_- \rho_{gg},
 \ee
 with 
 \bea
 R_{\pm} &=& |c|^4 G_C(\pm\omega_0) + |c|^4 g^2\alpha^2\omega_0^2 \left[\frac{G_H(\pm (\omega_0+\nu_0))}{\nu_0^2}\langle A_R^{\dag} (\pm\nu_0)A_R(\pm\nu_0)\rangle_{\rho_R^I} + G_C(\pm \omega_0)\sum_{\nu \in {\cal E}_R}\frac{T_C^{-1}}{\nu}\langle A_R^{\dag} (\nu)A_R(\nu)\rangle_{\rho_R^I} \right] \nn\\
 &&+{\cal O}\left({g^3\over |\nu|^3}\right).
 \eea
 As in the previous Section the analysis of the dynamics of $\langle A_R^{\dag} (\nu)A_R(\nu)\rangle_{\rho_R^I}$ reveals that its time derivative is of order $g^2/\nu^2$ implying that $\dot{R}_{\pm} = {\cal O}(g^4/\nu^4)$, justifying that we can safely take $R_{\pm}$ as constant. The dynamics of $\rho_{ee} = 1-\rho_{gg}$ is then
 \be
 \rho_{ee} = e^{-R t} \rho_{ee}(0) + \frac{1-e^{-R t}}{R} R_- =_{t\gg \tau_1} {R_-\over R},
 \ee
 where $R:=R_++R_-$. Substituting in \eqref{gexprhfsr}, \eqref{gexprhfs}, \eqref{gexprhfr} and using \eqref{Arhoeq} one obtains for the heat flows similar expressions as in the previous Section \ref{sectionho},
 \be
 \dot{Q}_{SR/C} = \omega_0{g^2\alpha^2\omega_0^2\over \nu_0^2}{G_C(\omega_0)G_H(\omega_0 + \nu_0) \over G_C(\omega_0)+G_C(-\omega_0)}  \left( e^{-{\omega_0\over T_C}}\langle A_R^{\dag}(\nu_0) A_R(\nu_0)\rangle_{\rho_R^I} -e^{-{\omega_0+\nu_0 \over T_H}}\langle A_R(\nu_0) A_R^{\dag}(\nu_0)\rangle_{\rho_R^I}\right) + {\cal O}(g^3/\nu_0^3),
 \ee
 \bea
 \dot{Q}_{SR/H}  &=&  -(\omega_0+\nu_0){g^2\alpha^2\omega_0^2\over \nu_0^2}{G_C(\omega_0)G_H(\omega_0 + \nu_0) \over G_C(\omega_0)+G_C(-\omega_0)}  \left( e^{-{\omega_0\over T_C}}\langle A_R^{\dag}(\nu_0) A_R(\nu_0)\rangle_{\rho_R^I} -e^{-{\omega_0+\nu_0 \over T_H}}\langle A_R(\nu_0) A_R^{\dag}(\nu_0)\rangle_{\rho_R^I}\right) + {\cal O}(g^3/\nu_0^3)\nn\\
 &=&- {\omega_0 +\nu_0 \over \omega_0} \dot{Q}_{SR/C} + {\cal O}\left({g^3\over \nu_0^3}\right), \label{tlsrelationhfxy}
 \eea
 \be\label{tlsqsx}
 \dot{Q}_{S/C} = -\dot{Q}_{S/H}+ {\cal O}\left({g^3\over \nu^3}\right) = \dot{Q}_{SR/C}+ {\cal O}\left({g^3\over \nu^3}\right).
 \ee
%
%
 %
Same identities as for harmonic oscillators (but the expression of the heat flows differs slightly, $G_C(\omega_0) +G_C(-\omega_0)$ in the denominator instead of $G_C(\omega_0)-G_C(-\omega_0)$). 
As in the previous Section, Eq. \eqref{tlsqsx} implies in particular that $\dot{E}_S =  {\cal O}\left({g^3\over \nu_0^3}\right)$. This is the usual condition of steady state for continuous thermal machines \cite{levy_quantum_2012,kosloff_quantum_2013, kosloff_quantum_2014, gelbwaser-klimovsky_heat-machine_2014} mentioned in the main text and valid for times $t$ much bigger than $\tau_{es}= [G_C(\omega_0)]^{-1}$.


  \section{Expression of $\dot{E}_R$}\label{appendixEr}
  The internal energy of $R$ is defined as (see also main text) $E_R := \bra H_R \ket_{\rho_{SR}}$. In term of $\rho_{SR}^I$, the density matrix of $SR$ in the interaction picture with respect to $H_{SR}$, the internal energy of $R$ can be re-written as 
  \be
  E_R = \bra H_{R}^I(t)\ket_{\rho_{SR}^I(t)},
  \ee 
  with $H_{R}^I(t) := e^{i t H_{SR}}H_R e^{-itH_{SR}} \ne H_R$. Then, the time derivative of $E_R$ is a sum of two contributions:
  \be
  \dot{E}_R = {\rm Tr} \dot{\rho}_{SR}^I(t) H^I_R(t) + {\rm Tr} \rho_{SR}^I(t) \dot{H}^I_R(t).
  \ee
  Using Eq. (23) of Methods `Why dispersive coupling?' (with $H_R^I(t)$ instead of $H_{SR}$) one can show straightforwardly that $ {\rm Tr} \dot{\rho}_{SR}^I(t) H^I_R(t) =0$. This is because in the interaction picture with respect to $H_{SR}$, $R$ does not interact with $S$ and therefore ``does not see" the baths. Then, only the second term contribute to $\dot{R}_R$. Up to second order in $g/|\nu|$ one can show that 
  \bea
    {\rm Tr} \rho_{SR}^I(t) \dot{H}^I_R(t) &=& ig \sum_{\nu \in {\cal E}_R} \nu e^{-i\nu t} \bra N_S A_R(\nu)\ket_{\rho_{SR}^I} +ig^2 \sum_{\nu,\nu' \in {\cal E}_R} \frac{\nu}{\nu'} \bra N_S^2  [A_R(\nu),A_R(\nu')]\ket_{\rho_{SR}^I} e^{-i\nu t} (e^{-i\nu't} -1)\nn\\
   &&  + {\cal O}\left(\frac{g^3}{|\nu|^3}\right).
    \eea
  The term of second order in $g/\nu$ is rapidly oscillating with $\nu \ne \nu'$, resulting in a contribution of higher order (after time-graining), and terms of $\nu'=-\nu$ sum up to zero since $\sum_{\nu \in {\cal E}_R} [A_R(\nu),A_R^{\dag}(\nu)] = 0$.
The first term contains the rapidly oscillating phase $e^{-i\nu t}$ so that the expression of $\bra N_S A_R(\nu)\ket_{\rho_{SR}^I} $ has to be derived retaining only terms oscillating at the frequency $\nu$,  becoming non-oscillating terms after multiplying by the phase $e^{-i\nu t}$. This can be done (for $S$ harmonic oscillator or two-level system) using Eq. (21) of Methods ``Why dispersive coupling?" (since Eq. (36) of Methods ``Expression of the baths dissipative operators" was obtained by neglecting fast oscillating terms).
 We finally find
  \be\label{gexprhfr}
 \dot{E}_R = - \sum_{\omega \in {\cal E}_S, \nu \in {\cal E}_R} \nu  G_H(\omega+\nu)\frac{g^{2}\alpha^2\omega^{2}}{\nu^2} \langle A^{\dag}_S(\omega)A_S(\omega)\rangle_{\rho_{S}^{eq}}\langle A_R^{\dag}(\nu)A_R(\nu)\rangle_{\rho_{R}^I} +{\cal O}\left({g^3\over|\nu|^3}\right).
 \ee
  Comparing with Eqs. \eqref{horelationhfxy} and \eqref{tlsrelationhfxy} when $S$ is a harmonic oscillator and a two-level system, respectively, one obtains the following fundamental relation, valid for $t\gg \tau_{es}$, 
\be\label{hodoter}
\dot{E}_R = {\nu_0 \over \omega_0 +\nu_0}\dot{Q}_{SR/H}+ {\cal O}\left({g^3\over \nu_0^3}\right) = - {\nu_0\over \omega_0}\dot{Q}_{SR/C}+ {\cal O}\left({g^3\over \nu_0^3}\right).
\ee

\section{Refrigeration conditions and efficiency}\label{main}
In Section \ref{appendixeval} we saw that the heat flows $\dot{Q}_{SR/j}$ takes almost the same expression for both two-level systems and harmonic oscillators. Thanks to that the following considerations are valid for both systems.
 From Section \ref{appendixeval} we have
\bea
\dot{Q}_{SR/C} &=& -{\omega_0\over\omega_0+\nu_0}\dot{Q}_{SR/H} \nn\\
&=&  \omega_0 {g^2\alpha^2\omega_0^2\over \nu_0^2}{G_C(\omega_0)G_H(\omega_0 + \nu_0) \over G_C(\omega_0)\pm G_C(-\omega_0)} \left( e^{-{\omega_0\over T_C}}\langle A_R^{\dag}(\nu_0) A_R(\nu_0)\rangle_{\rho_R^I} -e^{-{\omega_0+\nu_0 \over T_H}}\langle A_R(\nu_0) A_R^{\dag}(\nu_0)\rangle_{\rho_R^I}\right) + {\cal O}\left({g^3\over\nu_0^3}\right).\nn\\
\eea
The $\pm$ at the denominator corresponds to the possibility of $S$ being a harmonic oscillator or a two-level system.
The refrigeration condition corresponds to $\dot{Q}_{SR/C}\geq0$, which implies from \eqref{hodoter} $\dot{E}_R \leq0$, meaning that $R$ supplies energy to the refrigerator. The refrigeration condition is 
\be
{\omega_0+\nu_0\over T_H} -{\omega_0\over T_C} \geq \ln{{\langle A_R(\nu_0) A_R^{\dag}(\nu_0)\rangle_{\rho_R^I}\over \langle A_R^{\dag}(\nu_0) A_R(\nu_0)\rangle_{\rho_R^I}}}.
\ee
We defined the apparent temperature of $R$ as 
\be\label{gentemp}
{\cal T}_R := \nu_0 \left(\ln{{\langle A_R(\nu_0) A_R^{\dag}(\nu_0)\rangle_{\rho_R^I}\over \langle A_R^{\dag}(\nu_0) A_R(\nu_0)\rangle_{\rho_R^I}}}\right)^{-1}.
\ee
From \eqref{Arhoeq} (adapted to $R$) one can show that ${\cal T}_R = T_R$ when $R$ is in a thermal state at temperature $T_R$. More properties of the apparent temperature are mentioned in the main text. The refrigeration condition can be rewritten as
\be
\omega_0 \leq \nu_0 {T_C\over T_H-T_C} \left(1-{T_H\over {\cal T}_R}\right),
\ee  
which is the result announced and discussed in the main text. \\

The efficiency $\eta$ is defined as the ratio of the energy extracted from $C$, $\dot{Q}_{SR/C}$, by the energy invested by $R$, $-\dot{E}_R$, $\eta := {\dot{Q}_{SR/C} \over -\dot{E}_R}$. From the expressions \eqref{hodoter} we have
\be
\eta = {\omega_0 \over \nu_0} +  {\cal O}\left({g^3\over \nu_0^3}\right) \leq  {T_C\over T_H-T_C} \left(1-{T_H\over {\cal T}_R}\right),
\ee
and the upper bound is a direct consequence of the above refrigeration condition. This result is also announced and discussed in the main text.\\

Note that we defined the efficiency in terms of {\it rates} of the energy flows. Traditionally it is defined in terms of time-integrated flows of energy, i.e. finite difference of energy (between a time $t$ and the initial time). One can show that if we ignore the small time interval of order $\tau_1 =G_C^{-1}(\omega_0)$ before $S$ reaches the steady state, the efficiency $\eta_{int}:= {Q_{SR/C} \over -\Delta E_R}$ is also equal to $ {\omega_0 \over \nu_0} +  {\cal O}\left({g^3\over \nu_0^3}\right)$.

%
%
%
%
%

\section{Energy extraction conditions and efficiency}
We derive in this Section the conditions and the efficiency for the reverse operation of refrigeration: the storage in $R$ of energy extracted from the baths. Such storage operation is obtained by reversing the sign of all heat flows with respect to the refrigeration regime. 
The storage condition is $\dot{E}_R \geq 0$ which from \eqref{hodoter} implies $\dot{Q}_{SR/H} \geq 0$ and $\dot{Q}_{SR/C} \leq 0$. 
From Section \ref{appendixeval} we have
\bea
 \dot{E}_{R}  &=&  -\nu_0{g^2\alpha^2\omega_0^2\over \nu_0^2}{G_C(\omega_0)G_H(\omega_0 + \nu_0) \over G_C(\omega_0)+G_C(-\omega_0)}  \left( e^{-{\omega_0\over T_C}}\langle A_R^{\dag}(\nu_0) A_R(\nu_0)\rangle_{\rho_R^I} -e^{-{\omega_0+\nu_0 \over T_H}}\langle A_R(\nu_0) A_R^{\dag}(\nu_0)\rangle_{\rho_R^I}\right) + {\cal O}(g^3/\nu_0^3),\nn\\
 \eea
which imposes for the extraction condition,
\be
\omega_0 \geq \nu_0 {T_C\over T_H-T_C} \left(1-{T_H\over {\cal T}_R}\right),
\ee  
just the opposite of the refrigeration condition. The efficiency $\eta_e$ is defined as the energy stored in $R$, accounted by $\dot{E}_R$, divided by the cost in thermal energy from the hot bath, accounted by $\dot{Q}_{SR/H}$, $\eta_e :=\frac{\dot{E}_R}{\dot{Q}_{SR/H}}$. From \eqref{hodoter} we have that $\eta_e = \frac{\nu_0}{\omega_0+\nu_0} +{\cal O}(g^3/\nu_0^3)$. The storage extraction condition provides the upper bound,
\be
\eta_e \leq \left(1-\frac{T_C}{T_H}\right)\frac{{\cal T}_R}{{\cal T}_R - T_C},
\ee
assuming ${\cal T}_R \geq T_C$ (otherwise one can heat up $R$ trivially by thermal contact with $C$ or $H$). This is the expression mentioned and briefly discussed in the main text.

\section{Apparent temperature of non-degenerated systems} 
\subsection{Apparent temperature of squeezed states}
In this Section we consider that $R$ is a harmonic oscillator with the usual annihilation operator $A_R(\nu_0) = a$, creation operator $A_R^{\dag}(\nu_0) = a^{\dag}$, and the free Hamiltonian $H_R=\nu_0a^{\dag}a$, it is straight forward to show that ${\cal T}_R := \nu_0 \left[\ln{\left(1+\nu_0/ E_R\right)}\right]^{-1}$, which depends only on the average internal energy $E_R$.  Some consequences of such property are detailed in the main text. As a special case, we consider a squeezed thermal state of squeezing factor $r$ and thermal excitation corresponding to a temperature $T_R$. The mean energy of such squeezed state is \cite{Ferraro_2005} 
\be
E_R/\nu_0 = \sinh^2{r} + (\sinh^2r +\cosh^2r)\left(e^{\nu_0/T_R} -1\right)^{-1}.
\ee
The expression of the apparent temperature can be rewritten as
\bea
{\cal T}_R 
&=& \nu_0\left[\ln{\frac{ \tanh^2{r} + e^{\nu_0/T_R} }{ \tanh^2{r}e^{\nu_0/T_R} +1}}\right]^{-1},
\eea
which is the analogue of the expression used in \cite{huang_effects_2012,correa_quantum-enhanced_2014} for the effective temperature characterising the upper bound efficiency in presence of squeezed baths. \\

\subsection{Apparent temperature of a non-degenerated finite-level system}
Let's consider now a N-equidistant-level system. The ladder operator takes the form 
\be
A_R(\nu_0) = \sum_{n=1}^{N-1} c_{n,n+1} |n\rangle \langle n+1|,
\ee
where $|n\rangle$ is the eigenstate of the level $n$ and $c_{n,n+1} := \langle n|A_R|n+1\rangle$. The products of the upwards and downwards ladder operators give
\bea
A_R^{\dag}(\nu_0)A_R(\nu_0) &=& \sum_{n=1}^{N-1}|c_{n,n+1}|^2 |n+1\rangle \langle n+1|,\\
A_R(\nu_0)A_R^{\dag}(\nu_0) &=& \sum_{n=1}^{N-1}|c_{n,n+1}|^2 |n\rangle \langle n|.
\eea
From the definition of the apparent temperature \cite{latune_apparent_2018} provided in the main text, we obtain
\be
{\cal T}_R = \nu_0 \left( \log{\frac{\sum_{n=1}^{N-1} |c_{n,n+1}|^2 \rho_n}{\sum_{n=1}^{N-1}|c_{n,n+1}|^2\rho_{n+1}}}\right),
\ee
where $\rho_n:= \langle n| \rho_R^I |n\rangle$ is the population of the level $n$. For $N\geq3$ it appears that changing the value of the some $\rho_n$ can alter ${\cal T}_R$ but not necessarily the internal energy $E_R$, and reciprocally.  
It is more apparent if one assumes that the transition amplitudes $\langle n|A_R| n+1\rangle $ are independent of $n$ implying the following simple expression for the apparent temperature,
\be
{\cal T}_R = \nu_0 \left(\ln{\frac{1-\rho_N}{1-\rho_1}}\right)^{-1}.
\ee
Now we ask the question, given a fixed average energy $E_R$, what is the maximal achievable value of ${\cal T}_R$?
A simple way to derive the solution is by answering the reverse question: we look for the smallest $E_R$ compatible with a fixed apparent temperature ${\cal T}_R$. \\

We first assume that ${\cal T}_R$ is positive. This fixed the value of $\rho_N$ in term of $\rho_1$, $ \rho_N = 1-e^{\nu_0/{\cal T}_R}(1-\rho_1)$, with $\rho_1$ restricted to the range of values $1-e^{\nu_0/{\cal T}_R} \leq \rho_1\leq 1$ in order to $\rho_N$ be in the interval $[0;1]$. In case the population $\rho_1$ and $\rho_N$ do not sum up to 1, $\rho_1+\rho_N <1$, some other levels have to be populated. The choice which leads certainly to the lowest mean energy is populating the lowest available level, the level 2 (assuming $N\geq3$). Then, we choose $\rho_2 = 1- \rho_1-\rho_N$, with the allowed range of values for $\rho_1$:
\be
\frac{e^{\nu_0/{\cal T}_R} -1}{e^{\nu_0/{\cal T}_R}+1} \leq \rho_1 \leq \frac{e^{\nu_0/{\cal T}_R}}{e^{\nu_0/{\cal T}_R}+1}.
\ee
It follows that the double constraint $\rho_N$, $\rho_2$ $\in [0;1]$ implies $1-e^{-\nu_0/{\cal T}_R}\leq \rho_1 \leq \frac{e^{\nu_0/{\cal T}_R}}{e^{\nu_0/{\cal T}_R}+1}$. The mean energy of such a state is $E_R = \nu_0\rho_2 + \nu_0(N-1)\rho_N$, choosing a ground state energy equal to zero. Substituting $\rho_2$ and $\rho_N$ by their expression in term of $\rho_1$ we have 
\be
E_R = \nu_0\{N-2 + (\rho_1-1)[(N-2)e^{\nu_0/{\cal T}_R}-1] \}.
\ee
For $N\geq 3$, the minimal energy is $E_{R,min}({\cal T}_R) = \nu_0e^{-\nu_0/{\cal T}_R}$, achieved for the minimal allowed value $\rho_1=1-e^{-\nu_0/{\cal T}_R}$. Reversing the relation we obtained the maximal apparent temperature for a given energy $E_R \leq \nu_0$,
\be\label{trmax}
{\cal T}_{R,max} = \nu_0 [\ln{\nu_0/E_R}]^{-1}.
\ee
For $E_R \geq \nu_0$, the maximal apparent temperature is {\it negative}. Its expression can be derived in the same way as done above but assuming ${\cal T}_R$ negative.\\

Finally, a thermal state at temperature $T_R$ has an internal energy equal to
\be 
E_R^{th}(T_R) = \nu_0[(e^{\nu_0/T_R}-1)^{-1} - N(e^{N\nu_0/T_R}-1)^{-1}].
\ee
Then, thermal state of infinite temperature has an energy equal to $\nu_0(N-1)/2$. It follows that for $N\geq 4$, thermal states of {\it positive} temperatures can be manipulated with no energy change into non-thermal states of {\it negative} ${\cal T}_R$. 
 Substituting $E_R^{th}$ into \eqref{trmax} one can find numerically the maximum of ${\cal T}_{R,max}/T_R$.
 For instance, for $N=3$, we find ${\cal T}_{R,max}/T_R \rightarrow 1.5 $ (achieved for $T_R$ going to infinity). For $N\geq4$ the ratio ${\cal T}_{R,max}/T_R$ is indeed unbounded since a thermal energy equal to $\nu_0$ (corresponding to a finite positive temperature $T_R$) is enough to reach an infinite apparent temperature. \\

\end{document}